\documentclass[sigconf]{acmart}

\usepackage{amsmath}
\DeclareMathOperator{\softmax}{softmax}
\AtBeginDocument{}

\copyrightyear{2026}
\acmYear{2026}
\setcopyright{cc}
\setcctype{by}
\acmConference[RecSys '26]{20th ACM Conference on Recommender Systems}{September 27-October 02, 2026}{Minneapolis, MN, USA}
\acmBooktitle{20th ACM Conference on Recommender Systems (RecSys '26), September 27-October 02, 2026, Minneapolis, MN, USA}
\acmDOI{10.1145/3773078.3831864}
\acmISBN{979-8-4007-2284-4/2026/09}

\begin{document}

\title[SequenceO1]{SequenceO1: End-to-End Ultra-Long (100K) Sequence Modeling in Recommendation with Low-Rank Caching}

\author{Lin Guan}
\authornote{Contributed equally.}
\email{guanlin.13@gmail.com}
\orcid{0009-0006-6034-4819}
\affiliation{\institution{ByteDance}
  \city{Beijing}
  \country{China}
}

\author{Jia-Qi Yang}
\authornotemark[1]
\email{yangjiaqi.yjq@bytedance.com}
\orcid{0000-0002-6331-0829}
\affiliation{\institution{ByteDance}
  \city{Shanghai}
  \country{China}
}

\author{Zhishan Zhao}
\authornotemark[1]
\email{zhaozhishan@bytedance.com}
\orcid{0009-0000-4023-2197}
\affiliation{\institution{ByteDance}
  \city{Beijing}
  \country{China}
}

\author{Jiaqi Huang}
\authornotemark[1]
\email{jiaqihuang.0612@gmail.com}
\orcid{0009-0006-6937-3130}
\affiliation{\institution{ByteDance}
  \city{Beijing}
  \country{China}
}

\author{Hangyu Wang}
\email{wanghangyu.123@bytedance.com}
\orcid{0009-0006-0410-5426}
\affiliation{\institution{ByteDance}
  \city{Shanghai}
  \country{China}
}

\author{Longbin Li}
\email{mengxinghe@bytedance.com}
\orcid{0009-0006-7310-9412}
\affiliation{\institution{ByteDance}
  \city{Beijing}
  \country{China}
}

\author{Beichuan Zhang}
\email{zhangbeichuan.123@bytedance.com}
\orcid{0000-0003-1163-4036}
\affiliation{\institution{ByteDance}
  \city{Beijing}
  \country{China}
}

\author{Haonan Jiang}
\email{jianghaonan.1004@bytedance.com}
\orcid{0009-0005-6227-2057}
\affiliation{\institution{ByteDance}
  \city{Shanghai}
  \country{China}
}

\author{Jinan Ni}
\email{nijinan@bytedance.com}
\orcid{0009-0004-3287-4770}
\affiliation{\institution{ByteDance}
  \city{Shanghai}
  \country{China}
}

\author{Xiangyu Fan}
\email{xiangyufan.ptr@bytedance.com}
\orcid{0009-0004-3337-8728}
\affiliation{\institution{ByteDance}
  \city{Shanghai}
  \country{China}
}

\author{Xiaowen Li}
\email{lixiaowen.911@bytedance.com}
\orcid{0009-0000-9688-8339}
\affiliation{\institution{ByteDance}
  \city{Beijing}
  \country{China}
}

\author{Ziyao Ren}
\email{renziyao.99@bytedance.com}
\orcid{0009-0009-6330-1306}
\affiliation{\institution{ByteDance}
  \city{Beijing}
  \country{China}
}

\author{Yuhang Qi}
\email{qiyuhang@bytedance.com}
\orcid{0009-0008-7659-3160}
\affiliation{\institution{ByteDance}
  \city{Hangzhou}
  \state{Zhejiang}
  \country{China}
}

\author{Xiaolong Zhu}
\email{zhuxiaolong.auto@bytedance.com}
\orcid{0009-0004-0747-3721}
\affiliation{\institution{ByteDance}
  \city{Beijing}
  \country{China}
}

\author{Xuanyuan Luo}
\email{xuanyuanluo@bytedance.com}
\orcid{0009-0008-8412-3222}
\affiliation{\institution{ByteDance}
  \city{Hangzhou}
  \state{Zhejiang}
  \country{China}
}

\author{Qiwei Chen}
\authornote{Corresponding author.}
\email{chenqiwei05@gmail.com}
\orcid{0009-0002-8920-1716}
\affiliation{\institution{ByteDance}
  \city{Shanghai}
  \country{China}
}

\author{Yi Cheng}
\email{chengyi.23@bytedance.com}
\orcid{0009-0006-2950-1548}
\affiliation{\institution{ByteDance}
  \city{Beijing}
  \country{China}
}

\author{Lele Yu}
\email{yulele@bytedance.com}
\orcid{0009-0004-4199-861X}
\affiliation{\institution{ByteDance}
  \city{San Jose}
  \state{CA}
  \country{USA}
}

\renewcommand{\shortauthors}{Guan et al.}

\begin{abstract}
Modern short-video recommenders must exploit ultra-long user histories---which can reach hundreds of thousands or even millions of interactions per user---but are constrained by strict latency and training-throughput budgets.
At the 100K scale, the bottleneck is systemic, spanning feature storage, communication, and computation in both training and serving.
Existing solutions based on truncation, multi-stage retrieval, or length extrapolation either sacrifice end-to-end modeling or retain substantial length-dependent system cost.

We present \textbf{SequenceO1}, an end-to-end framework deployed at full traffic on Douyin at the \textbf{100K} scale and designed to extend to million-scale histories.
The name reflects its cache-hit path, whose cost is $O(1)$ with respect to the raw ultra-long sequence length once the fixed-size sketch is available.
At the model level, we propose \textbf{Sketch Attention (SA)}, which compresses an ultra-long history into a fixed-size, user-only sketch using learnable prototypes and \emph{prototype-wise} normalization (each token distributes mass over prototypes).
We then perform target-conditioned reasoning at two time scales: STCA over a recent 10K suffix for recency and STCA over the fixed-size sketch for ultra-long signals, followed by lightweight fusion.

At the system level, a training-side local key--value cache reuses user-only sketches across repeated instances of the same user, and the same cacheable state is reused across consecutive serving requests.
We further improve efficiency with multi-request user-level batching in training and a fused FlashSA kernel for sketching under ragged batching.
Together, these model and system optimizations make end-to-end 100K sequence modeling practical in production and provide a scalable path toward million-scale histories.
\end{abstract}



\maketitle

\section{Introduction}

\begin{figure}[t]
  \centering
  \includegraphics[width=\columnwidth]{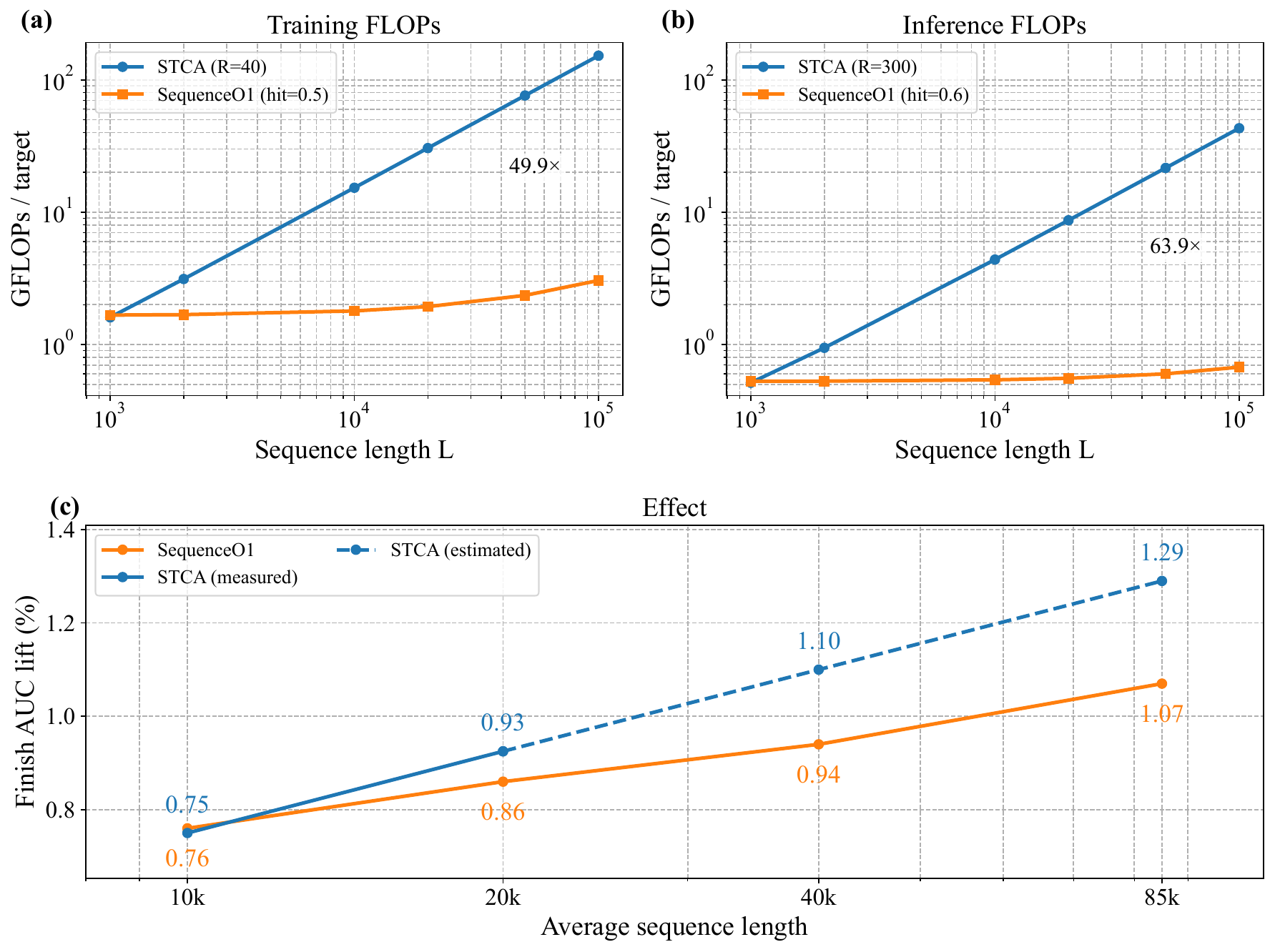}
\caption{\textbf{Compute-efficient scaling to 100K.}
(a) Training FLOPs of the ultra-long sequence branch for direct STCA and SequenceO1 from 1K to 100K under the observed training-side MRLB operating point ($R{=}40$, $p_{\mathrm{hit}}{=}0.5$); at 100K, SequenceO1 is $49.9\times$ cheaper than STCA.
(b) Inference FLOPs of the same branch under the observed serving-side reuse operating point ($R{=}300$, $p_{\mathrm{hit}}{=}0.6$); at 100K, SequenceO1 is $63.9\times$ cheaper than STCA. Common recent-10K and ranker-side costs are omitted from both.
(c) Finish AUC improvement over the STCA(512) baseline in the ablation setting versus average length; at 100K truncation (avg.\ 85K), SequenceO1 retains $83\%$ of directly scaled STCA's gain ($+1.07\%$ vs.\ $+1.29\%$).}
  \label{fig:complexity}
\end{figure}

\begin{figure*}[t]
  \centering
  \includegraphics[width=\textwidth]{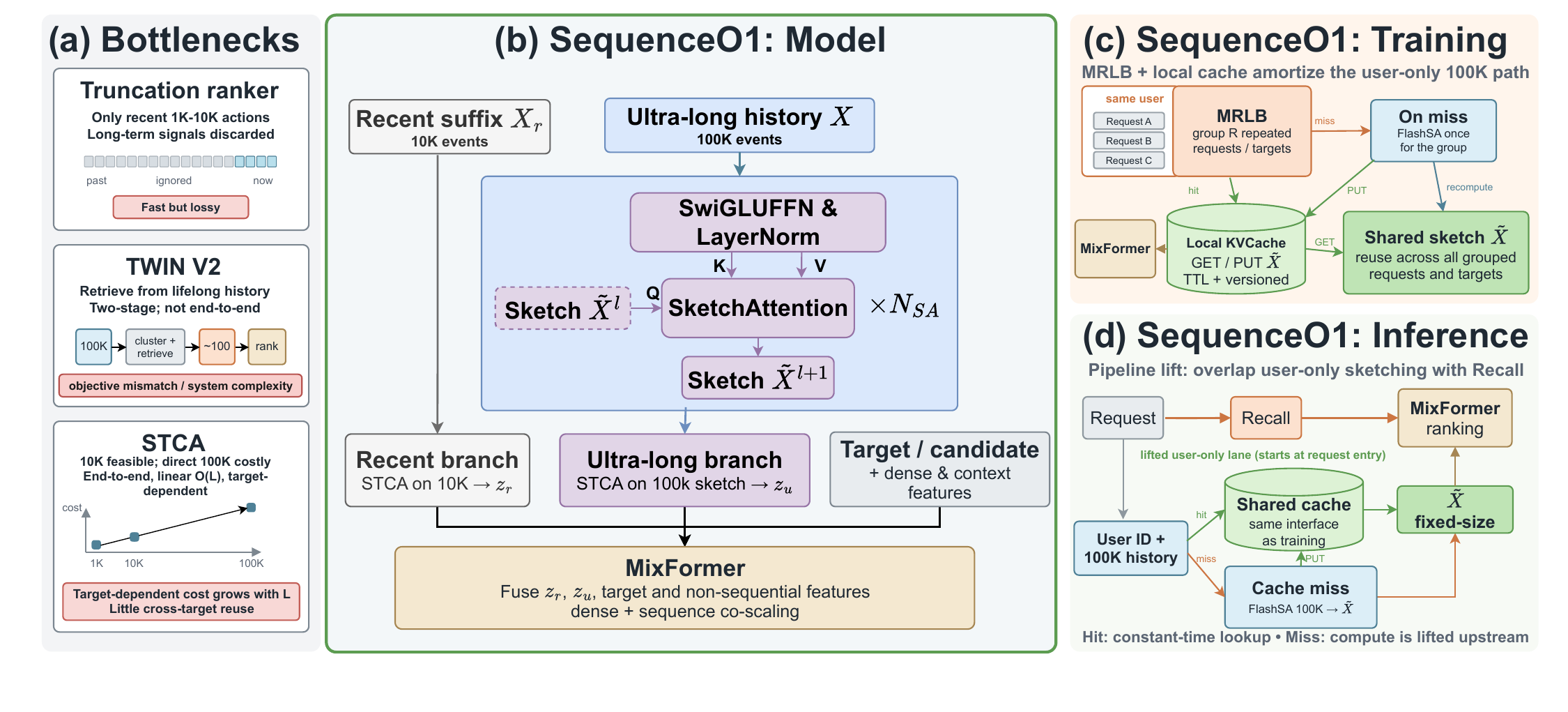}
  \caption{Overview of SequenceO1. (a) Existing approaches to the storage, communication, and computation bottlenecks of ultra-long sequence modeling---truncation, TWIN V2 two-stage retrieval, and direct STCA scaling---and their respective quality and system trade-offs. (b) Model architecture: stacked Sketch Attention (SA) compresses the 100K history into a fixed-size sketch, followed by recent-10K and ultra-long sketch branches and downstream MixFormer fusion. (c) Training: MRLB and the local KVCache amortize user-only sketching across grouped requests and targets, with FlashSA executed once per group on a cache miss. (d) Inference: the shared cache reuses fixed-size sketches, while pipeline lift overlaps miss-side FlashSA computation with recall.}
  \Description{SequenceO1 compresses a 100K user history into a fixed-size sketch through prototype-based Sketch Attention, refines the sketch, and combines recent-history and ultra-long-history branches in a downstream ranker. MRLB, a shared cache, pipeline lift, and FlashSA reduce repeated feature storage, communication, and user-only sketch computation during training and inference.}
  \label{fig:intro-overview}
\end{figure*}

Short-video recommendation at billion scale depends critically on modeling user behavioral histories.
Behavior-sequence modeling has been widely studied in modern recommender systems, from target-conditioned interest modeling to Transformer-based sequential recommendation~\cite{DBLP:conf/kdd/ZhouZSFZMYJLG18,DBLP:conf/aaai/ZhouMFPBZZG19,DBLP:conf/icdm/KangM18,DBLP:conf/cikm/SunLWPLOJ19,DBLP:conf/ijcai/0001LGQZ0T23}.
On Douyin, a single year of consumption can easily accumulate on the order of $10^5$ interactions per user.
Such ultra-long histories contain stable preferences, recurring intents, and periodic patterns that are often missed by short windows.
However, the fine-ranking stage operates under strict latency and training-throughput constraints, making it difficult to exploit these histories directly in an end-to-end manner~\cite{DBLP:conf/recsys/CovingtonAS16,DBLP:journals/corr/abs-1906-00091,DBLP:conf/isca/MudigereHHJT0LO22}.
At 100K, this is not merely a per-layer compute problem: embedding and sequence features enlarge training samples and cache footprints, moving them stresses host--device and distributed communication, and processing them increases both training and serving computation.

Production systems therefore often rely on truncation or multi-stage long-history pipelines that retrieve target-relevant behaviors, compress long histories, or combine both before final ranking~\cite{DBLP:conf/cikm/PiZZWRFZG20,DBLP:conf/sigir/Qin0WJF020,DBLP:conf/kdd/ChangZFZGLHLNSG23,DBLP:conf/cikm/CaoZFHXCC22}.
A representative baseline in our stack follows TWIN V2~\cite{DBLP:conf/cikm/SiGSZLHCYZLZZN024}, which uses offline hierarchical clustering to compress lifelong behaviors and online retrieval with cluster-aware target attention.
In our implementation, the clustered history contains about 10K entries, from which a General Search Unit (GSU) retrieves a small target-relevant subset for downstream fine-grained interest modeling.
Although effective, this design separates long-history compression and retrieval from the final ranking objective, limiting direct end-to-end optimization over the raw ultra-long history and adding system complexity~\cite{chen2021endtoenduserbehaviorretrieval,chen2022efficientlongsequentialuser}.

A more direct route is end-to-end long-sequence ranking with length extrapolation: train on shorter histories and serve on longer ones.
Meta's HSTU uses stochastic length sampling to reduce training cost, while Douyin's STCA adopts a train-sparsely/infer-densely regimen~\cite{DBLP:conf/icml/ZhaiLLWLCGGGHLS24,ding2026ultrahstu,DBLP:journals/corr/abs-2511-06077}.
The reported extrapolation ranges nevertheless remain coupled to training length: ULTRA-HSTU reports an average training length of about 4,400 for inference at 16,384, while STCA studies average training lengths of roughly 2--2.5K for serving at 10K---about four to five times longer in both cases.
Reaching 100K by extrapolation would therefore still require substantial training contexts, retaining high feature-sample storage, communication, and training-compute costs.
Nor does extrapolation remove the serving bottleneck: vanilla HSTU's self-attention is $O(L^2)$, whereas STCA reduces per-layer complexity to $O(L)$ but still incurs inference cost that grows linearly with the raw history length.
Moreover, STCA's target-conditioned cross attention is query-dependent and leaves little reusable computation across targets.
Thus, the 100K regime requires coordinated model and system design that avoids repeatedly storing, moving, and processing the raw ultra-long history, rather than optimizing per-layer computation alone~\cite{DBLP:journals/corr/abs-2006-04768,DBLP:conf/www/WuLGYYZY21}.

Another route is to decouple ultra-long user modeling from online ranking.
Recent multi-stage systems such as LLaTTE~\cite{xiong2026llattescalinglawsmultistage} and SOLARIS~\cite{liu2026solarisspeculativeoffloadinglatentbased} scale recommendation models by transferring asynchronous user representations to online rankers.
This reduces online computation, but it also introduces representation bottlenecks, objective mismatch, and freshness constraints between the upstream user model and the final ranking objective~\cite{chen2021endtoenduserbehaviorretrieval,chen2022efficientlongsequentialuser,DBLP:conf/recsys/0004ZZWZTZZWWC22}.
Therefore, a desirable 100K solution should reduce both training and serving cost while preserving an end-to-end optimization path for final ranking.

Our key observation is that the ultra-long portion of user history should be both \emph{compressible} and \emph{reusable}.
Prior work on efficient attention suggests that long sequences often admit compact representations along the length dimension~\cite{DBLP:journals/corr/abs-2006-04768}, and fixed-size latent summaries have also been effective for compressing large inputs before downstream reasoning~\cite{DBLP:conf/icml/LeeLKKCT19,DBLP:conf/icml/JaegleGBVZC21}.
In our setting, the 100K sequence should be summarized into a compact representation that depends only on user-side signals.
If such a fixed-size user representation can be computed once, cached, and reused across repeated training instances, targets, and consecutive requests, then the expensive target-conditioned module no longer needs to operate directly on the raw 100K history~\cite{zhou2024ercacheefficientreliablecaching}.

Motivated by this observation, we present \textbf{SequenceO1}, an end-to-end framework deployed at full traffic on Douyin that scales sequence ranking to the \textbf{100K} regime.
The name ``SequenceO1'' refers to the cache-hit path: once the fixed-size sketch is available, the raw 100K feature sequence need not be materialized, transferred, or processed for that sample, so the path is $O(1)$ with respect to the raw ultra-long sequence length.
As shown in Fig.~\ref{fig:intro-overview}, SequenceO1 combines a \emph{compress-then-reason} model design with \emph{cache-first} system amortization across training and inference.
At the model level, we introduce \textbf{Sketch Attention (SA)}, which compresses the full $n{=}100$K history into a fixed-size sketch $\widetilde{X}\in\mathbb{R}^{k\times d}$, with $k$ flexibly ranging from several hundred to several thousand.
SA uses learnable prototypes and \emph{prototype-wise} normalization so that each history token is allocated across prototypes, producing a target-agnostic user sketch.
This sketch can be viewed as an implicit low-rank representation along the length dimension, with size independent of $n$.
SequenceO1 then performs target-conditioned reasoning at two time scales: STCA over a recent 10K suffix for recency-sensitive signals, and STCA over the fixed-size sketch for ultra-long signals, followed by lightweight fusion.

At the system level, we first use a training-side local key--value cache (local KVCache) to reuse user-only sketches across repeated instances of the same user~\cite{zhou2024ercacheefficientreliablecaching}.
The same cacheable sketch is reused across consecutive serving requests; on cache hits, the system skips repeated raw-history feature storage and communication as well as the $n$-dependent sketch computation, and STCA reasoning is always performed on fixed-size inputs.
We further improve efficiency with multi-request user-level batching (MRLB), pipeline lift, and a fused FlashSA kernel inspired by IO-aware attention kernels~\cite{DBLP:conf/nips/DaoFERR22,dao2023flashattention2fasterattentionbetter}, which avoids materializing large sketching intermediates under ragged batching.
Unlike multi-stage representation-transfer systems, SequenceO1 keeps the ultra-long sequence path jointly optimized with the final ranker while making the 100K computation fixed-size, cacheable, and reusable.
We compare these scaling routes in \S\ref{sec:exp-transfer-ratio}.

Figure~\ref{fig:complexity} summarizes the resulting compute--quality trade-off.
Compared with na\"ively applying STCA at 100K, SequenceO1 is substantially cheaper under training-side local KVCache and MRLB as well as serving-side cache reuse, while retaining most of the quality gains from directly scaling sequence length.
This enables end-to-end sequence ranking to move from the 10K regime to 100K under real production constraints.

We summarize our contributions as follows:
\begin{itemize}
\item \textbf{End-to-end 100K ranking with SA+STCA.}
We extend end-to-end long-sequence ranking to \textbf{100K} histories by compressing ultra-long histories with Sketch Attention and applying target-conditioned STCA over both the sketch and a recent 10K suffix.

\item \textbf{A cacheable low-rank sketch for ultra-long histories.}
We design SA as a target-agnostic compression module that produces a fixed-size user-only sketch. The sketch is differentiable, length-agnostic, and reusable across targets and requests.

\item \textbf{Cache-first 100K training and serving.}
Training-side local KVCache and serving-side sketch reuse, complemented by MRLB, pipeline lift, and FlashSA, eliminate repeated $n$-dependent feature storage, communication, and computation on cache hits.
\end{itemize}

The remainder of the paper is organized as follows.
\S\ref{sec:background} reviews STCA and long-sequence compression, \S\ref{sec:method} presents the SequenceO1 model, \S\ref{sec:system} describes the system design, and \S\ref{sec:experiments} reports offline and online results on Douyin.

\section{Background and Notations}
\label{sec:background}

\subsection{Problem Setting}

We study the final-ranking (``fine-ranking'') stage of a large-scale short-video recommender system on Douyin. Given a request from user $u$, the ranker scores candidate videos $t$ under strict latency and cost constraints, using user/request features, item features, and the user interaction history $\mathcal{H}$ with supervision signals (e.g., finish, click, skip).

Our focus is on end-to-end modeling of ultra-long user histories in the fine-ranking stage. Following industrial practice, we truncate each user's history to a maximum length $n_{\max}=100\mathrm{K}$ for both training and serving, while preserving as much behavioral signal as possible within this budget.

\subsection{Notations}

We denote the user interaction history as
\begin{equation}
\mathcal{H}=\{(v_i,a_i)\}_{i=1}^{n},
\end{equation}
where $v_i$ is the $i$-th historical item and $a_i$ is the associated action type. Each event is embedded as $\mathbf{x}_i\in\mathbb{R}^{d}$, and the target item $t$ is represented by $\mathbf{x}_t\in\mathbb{R}^{d}$.

The history embedding matrix is
\begin{equation}
X=[\mathbf{x}_1,\ldots,\mathbf{x}_{n}]^\top \in \mathbb{R}^{n\times d}.
\end{equation}

We consider two time scales: an ultra-long history of length $n$ (up to $100\mathrm{K}$) and a recent suffix of length $L_r=10\mathrm{K}$.
We denote the recent suffix by
\begin{equation}
X_r = X_{n-L_r+1:n}\in\mathbb{R}^{L_r\times d},
\end{equation}
when $n\ge L_r$.
In our approach, the ultra-long history is compressed into a fixed-size sketch of length $k$ (typically $k\approx512$--1024), which is treated as constant with respect to $n$.

Throughout the paper, $n$ denotes the raw ultra-long history length, $L_r$ denotes the recent suffix length, and $k$ denotes the sketch length.
For a generic STCA module, we use $L$ to denote its input length; therefore $L=L_r$ for the recent branch and $L=k$ for the sketch branch.
We use $d$ for the behavior-embedding and SA width, $d_h$ for the STCA per-head width, and $D$ for the STCA model width after the sketch adapter when applicable.

\subsection{STCA Recap}
\label{sec:background-stca}

A central challenge in long-sequence ranking is the cost of modeling interactions over long histories. Standard self-attention incurs $O(L^2)$ complexity over a sequence of length $L$, which is prohibitive at industrial scales.

\emph{Stacked Target-to-History Cross Attention (STCA)} addresses this issue by removing history--history attention and instead applying stacked single-query cross attention from the target to the history. Given a target query $\mathbf{q}\in\mathbb{R}^{w}$ and history embeddings $X\in\mathbb{R}^{L\times w}$, one STCA layer computes
\begin{equation}
\mathrm{Attn}(\mathbf{q},X)=
\softmax\!\left(\frac{(\mathbf{q}W_Q)(XW_K)^\top}{\sqrt{d_h}}\right)\cdot (XW_V),
\end{equation}
where $W_Q,W_K,W_V\in\mathbb{R}^{w\times d_h}$.

With a single query, the per-layer cost scales linearly with $L$, enabling end-to-end ranking up to the 10K regime in production.
However, at $n{=}100$K, directly applying STCA is still too expensive for serving because its dominant target-to-history cross attention remains query-dependent and scales linearly with the raw history length.
This leaves little reusable computation across targets and motivates decoupling ultra-long user-side compression from target-conditioned reasoning.

\subsection{Length-Wise Compression Motivation}
\label{sec:background-linformer}

Efficient-attention studies suggest that long sequences admit substantial length-wise compression. Linformer motivates a short representation of length $k\ll n$ through low-rank attention approximation, while Set Transformer and Perceiver use inducing points or fixed latent arrays to summarize large inputs before expressive reasoning~\cite{DBLP:journals/corr/abs-2006-04768,DBLP:conf/icml/LeeLKKCT19,DBLP:conf/icml/JaegleGBVZC21}. Together, they show that a small set of learned summary tokens can trade sequence length for manageable computation.

Perceiver and poly-encoders typically use latent queries with token-wise softmax to \emph{select} inputs~\cite{DBLP:conf/icml/JaegleGBVZC21,DBLP:conf/iclr/HumeauSLW20}; SA instead applies prototype-wise softmax so every token is \emph{allocated} across prototypes, encouraging target-agnostic history coverage. Slot Attention~\cite{DBLP:conf/nips/LocatelloWUMHUD20} is closest in normalization spirit but targets iterative object-centric learning, whereas SA targets ultra-long ranking under strict latency constraints. Moreover, variable histories up to $n{=}100\mathrm{K}$ make explicit length-dependent projections inflexible, and generic compression does not naturally provide a fixed-size, end-to-end-compatible \emph{user-only} state reusable across targets and requests. These requirements motivate the length-agnostic, parameter-efficient, and cacheable SA mechanism introduced next.

\section{Method}
\label{sec:method}

We present an end-to-end framework for ultra-long sequence modeling at $n=100$K. The core idea is to \emph{compress} the ultra-long history into a compact, fixed-size representation and then perform target-conditioned \emph{reasoning} on this representation. Crucially, the compression depends only on user-side signals, making it reusable across multiple targets and consecutive requests.

\subsection{Sketch Attention (SA)}
\label{sec:method-sa}

Sketch Attention is named after the notion of a \emph{sketch}: a compact summary that preserves salient information from a much larger object under a fixed memory budget.
In our setting, the object is an ultra-long user history, and the sketch is a fixed-size set of user-side representation tokens.
This is analogous in spirit to streaming sketches such as Count Sketch~\cite{DBLP:conf/icalp/CharikarCF02}, but SA is learned end-to-end and optimized for downstream ranking rather than for recovering hand-designed statistics.

Given the ultra-long history embeddings $X\in\mathbb{R}^{n\times d}$, we learn $k$ trainable \emph{prototypes}, flexibly numbering from several hundred to several thousand,
\begin{equation}
P^{(0)} = [\mathbf{p}_1,\ldots,\mathbf{p}_k]^\top \in \mathbb{R}^{k\times d},
\end{equation}
which act as a compact set of summary slots. To enable reuse across targets and requests, the prototypes are shared globally, so the resulting representation depends only on the user history.

We compute prototype--token affinities as
\begin{equation}
S = \frac{(P^{(0)}W_Q)(XW_K)^\top}{\sqrt{d}} \in \mathbb{R}^{k\times n},
\end{equation}
where $W_Q, W_K \in \mathbb{R}^{d\times d}$.

\paragraph{Prototype-wise normalization.}
Instead of normalizing over tokens as in standard attention, we normalize over the \emph{prototype} dimension:
\begin{equation}
A = \softmax_{\text{proto}}(S)\in\mathbb{R}^{k\times n}, \qquad
\sum_{j=1}^{k} A_{j,i}=1.
\end{equation}
Equivalently,
\begin{equation}
A_{j,i} =
\frac{\exp(S_{j,i})}
{\sum_{j'=1}^{k}\exp(S_{j',i})},
\end{equation}
so $A_{j,i}$ defines a token-to-prototype allocation $p(j\mid i)$.
Compared with token-wise normalization, this allocation avoids early token selection and encourages coverage of the entire history before target conditioning, which is important when the compressed representation must support diverse targets in a reusable manner.

\paragraph{Sketch construction.}
We aggregate the history embeddings to obtain a compact sketch:
\begin{equation}
\widetilde{X} = A X \in \mathbb{R}^{k\times d}.\label{eq:sa-sketch}
\end{equation}
This operation is a weighted aggregation over history tokens rather than token-wise selection.
The resulting sketch is fixed-size, target-agnostic, and fully differentiable, making it suitable for caching and end-to-end training.

\subsection{SA as an Implicit Length-Wise Projection}
\label{sec:method-linformer}

Equation~\eqref{eq:sa-sketch} can be viewed as a data-dependent length-wise projection: the dynamically constructed $A\in\mathbb{R}^{k\times n}$ compresses the history from $n$ tokens to $k\ll n$ sketch tokens before target-conditioned reasoning. Unlike fixed projections tied to a maximum sequence length~\cite{DBLP:journals/corr/abs-2006-04768}, $A$ is generated from the input history and learnable prototypes, making SA length-agnostic in parameterization and jointly optimized with the ranking objective. Because this projection uses only user-side signals, its fixed-size, target-agnostic output $\widetilde{X}\in\mathbb{R}^{k\times d}$ can be materialized once and reused across targets and requests. This is the operational meaning of \emph{low-rank caching}: expensive target-conditioned reasoning runs on the cached compact length-wise representation rather than the raw ultra-long history.

\subsection{Stacked Refinement}
\label{sec:method-stack}

A single SA block produces a $k\times d$ sketch. To improve capacity, we apply a small number of refinement steps with residual connections. Starting from $\widetilde{X}^{(0)} = P^{(0)}$, we iteratively compute
\begin{align}
S^{(\ell)} &=
\frac{(\widetilde{X}^{(\ell)}W_Q^{(\ell)})(XW_K^{(\ell)})^\top}{\sqrt{d}}, \\
A^{(\ell)} &= \softmax_{\text{proto}}\!\left(S^{(\ell)}\right), \qquad
\sum_{j=1}^{k} A^{(\ell)}_{j,i}=1, \\
\widehat{X}^{(\ell+1)} &= A^{(\ell)}X, \\
\overline{X}^{(\ell+1)} &= \mathrm{LN}\!\left(\widetilde{X}^{(\ell)} + \widehat{X}^{(\ell+1)}\right), \\
\widetilde{X}^{(\ell+1)} &= \mathrm{LN}\!\left(\overline{X}^{(\ell+1)} + \mathrm{FFN}(\overline{X}^{(\ell+1)})\right).
\end{align}

The residual structure stabilizes optimization and preserves useful information from previous iterations, while the FFN increases representational capacity beyond linear aggregation. In practice, we find that a small number of iterations ($N_{\mathrm{sa}}{=}2$) is sufficient for modeling ultra-long histories.

\subsection{Two-Time-Scale Reasoning with STCA}
\label{sec:method-fusion}

We combine two complementary time scales: a recent suffix for fine-grained recency modeling and a compressed sketch for ultra-long signals.

\paragraph{Recent history.}
We model the most recent $L_r{=}10$K events directly using STCA, where $X_r\in\mathbb{R}^{L_r\times d}$:
\begin{equation}
\mathbf{z}_{r} = \mathrm{STCA}_{10k}(\mathbf{x}_t, X_r).
\end{equation}

\paragraph{Ultra-long history.}
Before sketch-side STCA, a width adapter $W_A$, distinct from the assignment matrix $A$, maps the SA sketch to the STCA width, while $\phi_t$ maps the target representation:
\begin{equation}
\widehat{X} = \widetilde{X}W_A\in\mathbb{R}^{k\times D},\qquad
\widehat{\mathbf{x}}_t=\phi_t(\mathbf{x}_t)\in\mathbb{R}^{D}.
\end{equation}
We then apply STCA over the adapted sketch:
\begin{equation}
\mathbf{z}_{u} = \mathrm{STCA}_{\text{sketch}}(\widehat{\mathbf{x}}_t, \widehat{X}).
\end{equation}
Since $\widehat{X}$ has fixed length $k$, the cost of this branch is independent of the original sequence length once the sketch is available.

\paragraph{Fusion.}
The recent and ultra-long representations are passed to the downstream ranking backbone, which may combine them with target and non-sequential features through a lightweight MLP, a gated fusion module, or its native fusion mechanism. Figure~\ref{fig:intro-overview} shows MixFormer~\cite{huang2026mixformer} as our deployed instantiation, but SequenceO1 does not require a particular fusion backbone.

\section{System}
\label{sec:system}

This section describes the system design that makes 100K history modeling deployable at full traffic on Douyin.
Our guiding principle is \emph{amortization}: the expensive ultra-long compression is user-only and should be computed once, reused across candidates, and reused across nearby requests whenever possible.
We implement this principle with a training-side \textbf{local KVCache} as the primary reuse mechanism, the same cache interface for serving, \textbf{MRLB}, \textbf{pipeline lift}, and \textbf{FlashSA}.

\subsection{Training-Side Local KVCache and Serving Reuse}
\label{sec:system-cache}

During training, we maintain a local key--value cache (local KVCache) for the ultra-long SA sketch $\widetilde{X}_u\in\mathbb{R}^{k\times d}$ because it is target-agnostic and depends only on user-side signals.
Request-specific features, including the recent suffix $X_{u,r}$ and candidate-side features, are not cached.
Optionally, one can also cache sketch-side linear projections that remain user-only under a fixed model version; in our complexity analysis, we use the conservative setting where the adapter output is not cached.

Each cache entry is keyed by user id, model version, history timestamp, and sketch configuration.
We use TTL-based invalidation with capacity eviction to bound memory and staleness.
In training, a 3-hour TTL within MRLB groups yields an effective hit rate of about $p_{\mathrm{hit}}^{\mathrm{train}}\approx0.5$.
In serving, the same cache interface with a 1-hour TTL gives an empirical hit rate of about $p_{\mathrm{hit}}^{\mathrm{infer}}\approx0.6$.
This shared get/put interface reduces train/serve skew while substantially reducing repeated 100K feature storage, communication, and sketch computation.
The reported FLOP reductions use these observed hit rates as operating points.
On a cache miss, the additional raw-history computation is only the user-only SA sketching step, which can still be amortized by MRLB or executed upstream.

\subsection{Multi-Request Level Batching (MRLB)}
\label{sec:system-mrlb}

Even with SA, constructing $\widetilde{X}_u=\mathrm{SA}(X_u)$ over 100K tokens is expensive.
In production, a user often issues multiple consecutive requests within a short time window, while the long-term history changes slowly.
MRLB groups such requests from the same user and computes the ultra-long sketch once, reusing it for all targets in the group:
\begin{equation}
\widetilde{X}_u=\mathrm{SA}(X_u)\in\mathbb{R}^{k\times d}.
\end{equation}
Request-specific components, such as the recent suffix and candidate features, are still computed per request.

In practice, we cap the aggregation degree to keep memory stable and apply request-aware masking in the recent-history branch to prevent cross-request leakage.
By amortizing user-only sketching, MRLB reduces repeated data movement and improves training utilization in the 100K setting.

\subsection{Pipeline Lift}
\label{sec:system-lift}

Fine-ranking typically receives only a fraction of the end-to-end latency budget because recall, coarse ranking, filtering, and feature fetching already consume substantial time.
In conventional pipelines, ranking-side sequence modeling must wait until the candidate set is available.

Since $\widetilde{X}_u=\mathrm{SA}(X_u)$ depends only on user-side information, we can lift this computation to request entry or other upstream stages and pass the sketch forward as a user feature, as shown in Fig.~\ref{fig:intro-overview}.
On cache hits, a constant-time lookup supplies the fixed-size sketch before the raw 100K feature sequence is materialized and transferred to the model; on misses, the user-only computation can still run in parallel with upstream retrieval.
Thus, cache hits remove raw-length feature storage, communication, and computation from the sample path, while pipeline lift further reduces the latency impact on misses.

\subsection{FlashSA: Kernel Optimization for SA}
\label{sec:system-FlashSA}

A naive SA implementation materializes the prototype--token affinity matrix $S\in\mathbb{R}^{k\times n}$ and often the assignment matrix $A\in\mathbb{R}^{k\times n}$ from \S\ref{sec:method-sa}, leading to $O(kn)$ intermediate storage and heavy memory traffic at $n{=}100$K, plus low efficiency from multiple kernel launches under ragged batching.
We implement \textbf{FlashSA}, a fused kernel that streams the computation: it performs block-wise affinity computation, prototype-wise softmax statistics (max and log-sum-exp), and the final aggregation into $\widetilde{X}$ without storing the full $k\times n$ scores in HBM.
This reduces peak memory, improves throughput via fewer launches and better locality, and stabilizes mixed-precision execution through numerically robust softmax and accumulation.

Together, these components make the 100K sketch path reusable, cache-friendly, and hardware-efficient, enabling end-to-end 100K sequence modeling at billion scale on Douyin.

\section{Experiments}\label{sec:experiments}

\paragraph{Evaluation settings.}
We report two offline evaluation settings.
Most ablations use a lightweight setting with a simplified dense component for efficient experimentation, with STCA(512) as the baseline, and focus on Finish AUC.
The SA-vs-vanilla diagnostic additionally reports multi-task UAUC to verify that the normalization change improves several engagement objectives.
The production offline evaluation and online A/B test use the full production ranker: the baseline is STCA 10K with lifelong TWIN V2, while SequenceO1 removes TWIN V2 and replaces it with the end-to-end 100K sketch branch.

\subsection{Sketch Attention vs.\ Vanilla Attention}
\label{sec:exp-sa-vs-vanilla}

\begin{figure}[t]
  \centering
  \includegraphics[width=0.7\columnwidth]{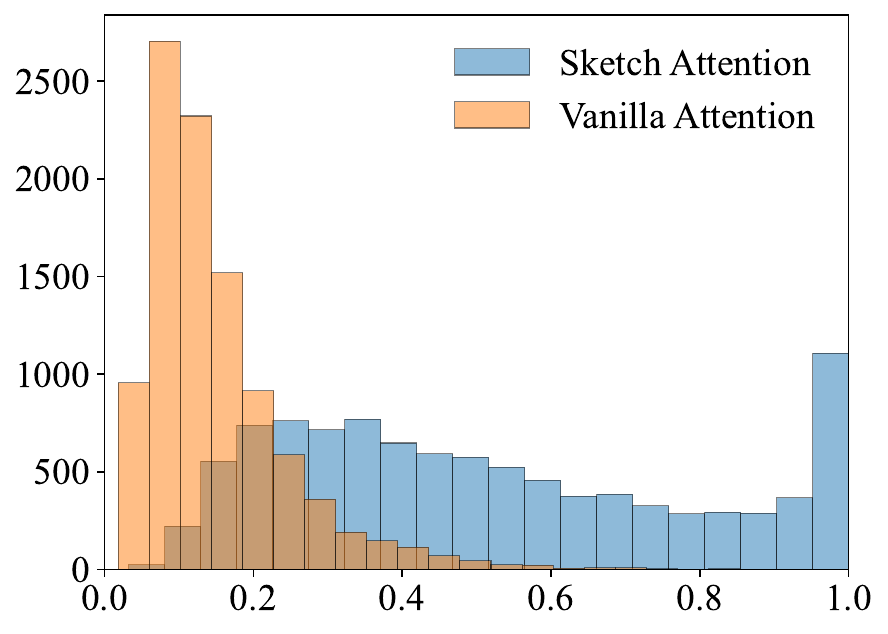}
  \caption{Distribution of per-token maximum normalized weights under the two normalization rules in the 100K setting; their scales differ because the normalization axes differ.}
  \label{fig:attn-score-dist}
\end{figure}

\begin{table}[t]
\centering
\caption{UAUC improvements of SA over vanilla attention under the 100K setting.}
\label{tab:uauc-sa-vs-vanilla}
\small
\resizebox{\columnwidth}{!}{\begin{tabular}{l c c c c c c}
\toprule
Task & like & follow & clicmmt & comment & share & finish \\
\midrule
$\Delta$UAUC & +0.10\% & +0.20\% & +0.09\% & +0.19\% & +0.14\% & +0.01\% \\
\bottomrule
\end{tabular}}
\end{table}

We analyze how prototype-wise normalization in Sketch Attention (SA) differs from vanilla attention under the 100K ablation setting.
We keep the same STCA(512)-based ranking setup and replace only the normalization mechanism in the ultra-long compression branch.
For each sampled example, we compute the prototype--token weight matrix $A$ and record the maximum normalized weight associated with every token.
Because prototype-wise and token-wise normalization operate over different axes, these maxima are not calibrated confidence scores and should not be compared as direct evidence of prototype specialization.
Figure~\ref{fig:attn-score-dist} instead reports the raw distributions under the respective conservation constraints as a descriptive diagnostic; the shift between the histograms reflects both the induced allocation behavior and the different normalization-domain sizes.

With prototype-wise normalization, each behavior token is explicitly allocated across prototypes, which is aligned with constructing target-agnostic summaries before target conditioning.
Such allocation may provide an interface for possible downstream sketch-to-behavior retrieval or localized refinement, although these operations are not evaluated components in this work.
The task-level evidence comes from Table~\ref{tab:uauc-sa-vs-vanilla}: prototype-wise normalization improves UAUC across multiple engagement objectives, supporting it as a better mechanism for constructing reusable 100K sketches in our setting.

\subsection{Ablations and Compression Methods}
\label{sec:exp-sa-ablation}

\begin{table}[t]
\centering
\caption{SA architecture ablations under the 100K ablation setting.
The final SA setting achieves +1.07\% Finish AUC over STCA(512); all numbers report changes relative to this final SA setting.}
\label{tab:sa-arch-ablation}
\small
\setlength{\tabcolsep}{4pt}
\renewcommand{\arraystretch}{1.05}
\begin{tabular}{lc}
\toprule
Variant & Change vs. final SA \\
\midrule
Final SA ($k{=}1\mathrm{K}$, $N_{\mathrm{sa}}{=}2$, $d{=}128$) & 0.00\% \\
$k$: 1K$\rightarrow$512 & -0.04\% \\
$k$: 1K$\rightarrow$2K & +0.03\% \\
$N_{\mathrm{sa}}$: 2$\rightarrow$3 & +0.02\% \\
$d$: 128$\rightarrow$64 & -0.10\% \\
w/o action-side SwiGLU fusion & -0.20\% \\
w/o per-layer Add\&Norm + SwiGLU FFN & -0.10\% \\
\bottomrule
\end{tabular}
\end{table}

\begin{table}[t]
\centering
\caption{Comparison of 100K compression methods in the ablation setting.
The base model uses STCA(512) only, while all non-base variants add an ultra-long compression branch to the same ranking setup.
For fair comparison, each compression method summarizes the 100K history into a compact $1K\times128$ representation.
Numbers report Finish AUC improvements over the STCA(512) base model.}
\label{tab:compression-methods}
\small
\setlength{\tabcolsep}{4pt}
\renewcommand{\arraystretch}{1.05}
\begin{tabular}{lc}
\toprule
Compression method & Gain vs. STCA(512) \\
\midrule
Base (STCA(512)) & 0.00\% \\
SA (final) & +1.07\% \\
TWIN V2 KMeans Clustering~\cite{DBLP:conf/cikm/SiGSZLHCYZLZZN024} & +0.30\% \\
Chunk mean pooling & +0.46\% \\
Position-bucket queries & +0.72\% \\
Lightning-Attention comp.~\cite{minimax2025minimax01scalingfoundationmodels} & +0.83\% \\
Recent-behavior query init.~\cite{DBLP:conf/recsys/ChaiRXYHZCLZYXR25} & +0.78\% \\
\bottomrule
\end{tabular}
\end{table}

Under the same 100K lightweight STCA(512) setup, Table~\ref{tab:sa-arch-ablation} reports changes relative to the final SA configuration, while Table~\ref{tab:compression-methods} reports absolute Finish AUC gains over STCA(512). The final setting---$k{=}1\mathrm{K}$, $N_{\mathrm{sa}}{=}2$, $d{=}128$, action-side SwiGLU fusion of item and action-type embeddings, and per-layer Add\&Norm with a SwiGLU FFN---achieves +1.07\%. Increasing to 2K prototypes or 3 layers adds only +0.03\% or +0.02\%, respectively; we treat these as cost--quality guidance rather than standalone significance evidence and retain the smaller production configuration. Reducing width or removing refinement degrades performance, with the -0.20\% from removing action-side fusion highlighting the value of action-aware sketching.

Table~\ref{tab:compression-methods} compares methods under the same $1\mathrm{K}\times128$ budget: K-means clustering inspired by TWIN V2~\cite{DBLP:conf/cikm/SiGSZLHCYZLZZN024}, neighboring-behavior chunk means, position-bucket queries over roughly 100 behaviors each, a Lightning-Attention long-sequence compressor, and queries initialized by the most recent 1K behaviors. SA performs best, showing that its gain comes from compression quality rather than representation size; its advantage over position buckets also suggests that recommendation histories are less locally structured than language because related behaviors can be far apart.

\subsection{Offline Performance on Douyin Dataset}
\label{sec:exp-offline}

We evaluate SequenceO1 on the full production Douyin offline dataset against \textbf{STCA 10K} with lifelong TWIN V2. SequenceO1 removes TWIN V2 in favor of the end-to-end 100K sketch branch, so this strict comparison requires it to improve ranking while absorbing the previous module's long-term signals. Table~\ref{tab:offline_metrics} reports relative AUC and UAUC improvements across engagement objectives.

\begin{table*}[t]
\centering
\caption{Offline improvements in the full production setting on the Douyin dataset.
The baseline is production STCA 10K with lifelong TWIN V2; SequenceO1 removes TWIN V2 and uses the end-to-end 100K sketch branch instead.}
\label{tab:offline_metrics}
\small
\begin{tabular}{lcccclccccc}
\toprule
Metric & Finish & Skip & Head & Like & Follow & Comment & Clicmmt & Share & Favourite & Dislike \\
\midrule
$\Delta$AUC  & +0.29\% & +0.23\% & +0.18\% & +0.09\% & +0.05\% & +0.04\% & +0.16\% & +0.18\% & +0.30\% & +1.29\% \\
$\Delta$UAUC & +0.40\% & +0.43\% & +0.46\% & +0.72\% & +0.58\% & +0.43\% & +0.61\% & +0.94\% & +1.47\% & +3.63\% \\
\bottomrule
\end{tabular}\end{table*}

SequenceO1 improves every evaluated objective despite removing TWIN V2, showing that the 100K branch successfully replaces the two-stage lifelong module while further improving all evaluated metrics in the full ranker. UAUC gains are strongest on preference-sensitive objectives (Like, Share, Favourite, and Dislike), while core consumption objectives (Finish and Skip) also improve, indicating more effective long-term personalization than the previous cluster-based pipeline.

\subsection{Online Performance}
\label{sec:exp-online-perf}
We deployed \textbf{SequenceO1} for one month on Douyin and Douyin Lite, replacing the lifelong TWIN V2 module in the production STCA 10K baseline with the end-to-end 100K sketch branch. Table~\ref{tab:online_ab} reports relative changes over control in 30-Day Activeness, Duration, Finish, Comment, Like, and Dislike, overall and by user activity.

\begin{table*}[h]
\caption{Online A/B results over the production STCA 10K + TWIN V2 baseline on Douyin and Douyin Lite (all statistically significant).}
\resizebox{\textwidth}{!}{
\renewcommand\arraystretch{1.08}
\begin{tabular}{lcccccccccccc}
\hline
                       & \multicolumn{6}{c}{\textbf{Douyin}}                                                          & \multicolumn{5}{c}{\textbf{Douyin Lite}}                                  &                  \\ \cmidrule(lr){2-7}\cmidrule{8-13}
                       & \textbf{30-Day Act.}$\uparrow$ & \textbf{Duration}$\uparrow$ & \textbf{Finish}$\uparrow$ & \textbf{Comment}$\uparrow$ & \textbf{Like}$\uparrow$ & \textbf{Dislike}$\downarrow$ & \textbf{30-Day Act.}$\uparrow$ & \textbf{Duration}$\uparrow$ & \textbf{Finish}$\uparrow$ & \textbf{Comment}$\uparrow$ & \textbf{Like}$\uparrow$ & \textbf{Dislike}$\downarrow$ \\ \hline

\textbf{Overall}       & +0.1968\%           & +1.4999\%         & +2.3256\%     & +3.5910\%       & +2.3617\%      & -6.9829\%  & +0.2335\%           & +1.6655\%          & +3.4872\%     & +8.6132\%        & +2.9399\%    & -5.9972\%    \\ \hline

\textbf{Low-active}    & +0.5484\%           & +2.1476\%         & +3.2040\%     & +2.9099\%       & +3.5739\%    &  -8.3061\%    & +0.6614\%           & +1.9989\%         & +3.9845\%      & +13.4620\%       & +6.1890\%   & -2.4977\%     \\

\textbf{Middle-active} & +0.5235\%           & +2.2408\%         & +3.1550\%     & +5.1155\%       & +3.0477\%    & -12.1796\%    & +0.4306\%           & +1.8468\%         & +3.0968\%      & +8.7780\%         & +0.0254\%    & -3.6990\%    \\

\textbf{High-active}   & +0.1789\%           & +1.6334\%         & +2.4733\%      & +3.5232\%       & +2.1611\%    &  -5.7208\%     & +0.3090\%           & +1.8161\%         & +3.4367\%      & +8.3048\%        & +3.2317\%  & -3.1752\%
\\

\textbf{Full-active}   & +0.0622\%           & +1.2705\%         & +2.0744\%      & +3.1494\%       & +2.1376\%    &  -7.8567\%     & +0.0853\%           & +1.5254\%         & +3.2774\%      & +7.5138\%        & +2.7786\%  & -6.5438\%
\\ \hline
\end{tabular}
}
\label{tab:online_ab}
\end{table*}

SequenceO1 significantly improves Activeness, Duration, Finish, Comment, and Like while reducing Dislike on both apps; Finish rises by +2.33\% on Douyin and +3.49\% on Douyin Lite, indicating better sustained consumption. Gains are stable across activity segments: low-active users improve in activeness and duration, suggesting re-engagement, while high-active users improve in Finish and Like. Overall, the model complements short-term signals with stable long-term preferences across platforms and cohorts.

\subsection{Relation to Two-Stage Transfer}
\label{sec:exp-transfer-ratio}

Recent industrial systems often use two-stage representation transfer: LLaTTE~\cite{xiong2026llattescalinglawsmultistage} reports a $50\%$--$53\%$ ratio from upstream user-modeling to downstream ranking gains, while SOLARIS~\cite{liu2026solarisspeculativeoffloadinglatentbased} reports about $42\%$ and $44\%$ on Instagram and Facebook products. Because these results come from different systems, they contextualize transfer efficiency rather than provide an apples-to-apples comparison; our same-system evidence comes from replacing production TWIN V2 with an end-to-end 100K sketch branch.

In this ablation, Fig.~\ref{fig:complexity}(c) shows that SequenceO1 retains $83\%$ of directly scaled STCA's gain at 100K truncation (85K average length): $+1.07\%$ versus $+1.29\%$ Finish AUC. Thus, it translates sequence-length scaling into ranking gains while preserving a cacheable path. Unlike two-stage transfer, its simpler sketch remains jointly optimized with the final ranker instead of becoming a fixed external feature, and could support future in-graph sketch retrieval or localized refinement, neither of which is evaluated in the current deployment.

\section{Related Work and Discussion}\label{sec:related}

\subsection{Modeling User Behavior Sequences}

User behavior modeling is central to recommender systems.
Early methods include item-to-item collaborative filtering and Markov transition models~\cite{DBLP:journals/internet/LindenSY03,DBLP:conf/www/RendleFS10}, while deep learning introduced session-based RNNs and large-scale industrial ranking architectures~\cite{DBLP:journals/corr/HidasiKBT15,DBLP:conf/recsys/CovingtonAS16}.
Attention-based models later became dominant: DIN and DIEN perform target-conditioned user-interest modeling~\cite{DBLP:conf/kdd/ZhouZSFZMYJLG18,DBLP:conf/aaai/ZhouMFPBZZG19}, SASRec and BERT4Rec use self-attention for sequential modeling~\cite{DBLP:conf/icdm/KangM18,DBLP:conf/cikm/SunLWPLOJ19}, and BST demonstrates Transformer-based behavior modeling in industrial ranking~\cite{DBLP:journals/corr/abs-1905-06874,huang2026mixformer}.
Multi-interest methods further capture diverse user intents at scale~\cite{DBLP:conf/cikm/LiLWXZHKCLL19}.

Recent production systems increasingly emphasize longer histories under serving constraints.
PinnerFormer and TransAct improve long user representation and real-time action modeling at Pinterest~\cite{DBLP:conf/kdd/PanchaZLR22,DBLP:conf/kdd/XiaEPBWGJFZZ23,DBLP:journals/corr/abs-2506-02267}.
Other systems retrieve, sample, or compress target-relevant subsets from lifelong behavior logs, including SIM, UBR4CTR, SDIM, TWIN, and TWIN V2~\cite{DBLP:conf/cikm/PiZZWRFZG20,DBLP:conf/sigir/Qin0WJF020,DBLP:conf/cikm/CaoZFHXCC22,DBLP:conf/kdd/ChangZFZGLHLNSG23,DBLP:conf/cikm/SiGSZLHCYZLZZN024}.
In particular, TWIN V2 uses offline hierarchical clustering to compress lifecycle behaviors into manageable clusters, followed by online retrieval and cluster-aware target attention for CTR prediction.
These methods are effective under serving constraints, but their retrieval or compression stages are typically separated from the final ranking objective.

\subsection{Caching and Reuse}
\label{sec:related-cache}
Caching strategies differ mainly by whether the cached object is end-to-end over raw histories and whether its size scales with sequence length.

UIC+MIMN~\cite{DBLP:conf/kdd/PiBZZG19} and HPMN~\cite{DBLP:conf/sigir/RenQF0ZBZXYZG19} maintain incrementally updated user states, enabling constant-time reads but not end-to-end optimization over the raw ultra-long sequence at inference time. CTR-oriented methods cache candidate-agnostic summaries through grouping, quantization, or clustering, including DGIN~\cite{DBLP:journals/corr/abs-2311-10764}, DMQN~\cite{DBLP:conf/sigir/WeiLX25}, and C-Former~\cite{DBLP:conf/cikm/00010LDLLZ25}; generative recommenders such as VISTA~\cite{DBLP:journals/corr/abs-2510-22049} cache summary tokens. These methods usually target shorter horizons or rely on separate pipelines. Transformer KV/prefix caching reduces recomputation but stores per-position states, so cache footprint grows with length~\cite{wang2026relaygrscalinglongsequencegenerative,li2026collectivekvdecouplingsharingcollaborative,DBLP:journals/corr/abs-2510-26104,DBLP:conf/kdd/0002000SHC25}. In contrast, SequenceO1 caches a \emph{fixed-size} user-only SA sketch ($100\mathrm{K}\!\rightarrow\!k$) in training and serving, making cache size independent of the original history length and eliminating repeated raw-length feature storage, communication, and model computation on cache hits.

SequenceO1 differs from multi-stage representation-transfer systems such as LLaTTE~\cite{xiong2026llattescalinglawsmultistage} and SOLARIS~\cite{liu2026solarisspeculativeoffloadinglatentbased} by keeping the ultra-long sequence path end-to-end optimized with the final ranking objective.
It uses a simpler cacheable sketch rather than a separate upstream user model, providing a practical alternative to the previous TWIN V2-style lifelong module while preserving a direct optimization path for final ranking.

\section{Conclusion}

SequenceO1 scales end-to-end sequence-based recommendation to 100K in production and supports extension toward million-scale histories by combining Sketch Attention, two-time-scale STCA reasoning, and training-first cache amortization. Douyin experiments show consistent gains in lightweight ablations, full production offline evaluation, and online A/B tests: a compact reusable sketch preserves most of the benefit of direct long-sequence scaling while substantially reducing feature storage, communication, and computation in training and serving. Future directions include hierarchical or multi-resolution sketches and adaptive capacity across users or behavior patterns; more broadly, the results show that coordinated model and system design can jointly deliver ultra-long modeling and production efficiency.

\appendix
\onecolumn

\section{Connection to Length-Wise Low-Rank Projection}
\label{app:linformer}

This section clarifies the connection between Sketch Attention (SA) and \emph{length-wise low-rank projection}.
Our goal is not to claim that SA is a strict instantiation of Linformer or that it inherits a specific approximation guarantee for self-attention.
Rather, we use the low-rank perspective as a conceptual lens: it helps explain why an ultra-long history of length $n$ may be summarized by a much shorter representation of length $k \ll n$ while retaining predictive signals for downstream ranking.

The key point is that SequenceO1 compresses the history \emph{along the length dimension}.
From this viewpoint, SA can be interpreted as constructing a data-dependent map from the original sequence of length $n$ to a fixed-size representation of length $k$, after which target-conditioned reasoning is performed on the compressed representation.
This perspective is closely related in spirit to prior low-rank efficient-attention methods, but differs in both construction and system role.

\subsection{Self-Attention as a Length-Wise Mapping}

Consider a single-head self-attention layer with query, key, and value matrices
\[
Q,K,V \in \mathbb{R}^{n\times d_h}.
\]
The attention output can be written as
\begin{equation}
\mathrm{SelfAttn}(Q,K,V) = \Pi V,
\qquad
\Pi = \softmax\!\left(\frac{QK^\top}{\sqrt{d_h}}\right)\in\mathbb{R}^{n\times n},
\end{equation}
where $\Pi$ is the attention weight matrix along the \emph{length dimension}.
Thus, self-attention can be interpreted as applying a length-wise transformation $\Pi$ to the value matrix $V$.

In general, $\Pi$ is dense and may have rank up to $n$.
Materializing and applying this $n\times n$ matrix incurs quadratic complexity in sequence length.
This observation motivates a natural question: does the effective interaction really require full rank along the length axis, or can it be well-approximated in a much lower-dimensional subspace?

\subsection{Low-Rank Approximation along the Length Dimension}

Linformer~\cite{DBLP:journals/corr/abs-2006-04768} proposes that, under certain assumptions, the attention matrix can often be well-approximated by a low-rank factorization along the length dimension.
Concretely, instead of operating on full-length keys and values, one projects them to a lower-dimensional subspace:
\begin{equation}
K' = E^\top K,\qquad V' = F^\top V,
\qquad E,F\in\mathbb{R}^{n\times k}, \quad k \ll n,
\end{equation}
so that attention is computed over $k$ positions rather than $n$.
The resulting approximation can be written as
\begin{equation}
\mathrm{SelfAttn}(Q,K,V) \approx
\softmax\!\left(\frac{QK'^\top}{\sqrt{d_h}}\right)V'.
\end{equation}

This formulation suggests that the effective dimensionality of sequence interaction along the length axis can be much smaller than $n$, and that a fixed projection size $k$ may suffice even as $n$ grows.
For our setting, this is an appealing intuition: ultra-long user histories may contain substantial redundancy, so a compact summary along the length dimension may preserve most of the useful information for ranking.

\subsection{Limitations of Explicit Length-Wise Projections}

While the above formulation is appealing, directly applying explicit length-wise projections in large-scale recommender systems is challenging:

\begin{itemize}
\item \textbf{Parameter scaling with maximum length.}
The projection matrices $E,F\in\mathbb{R}^{n\times k}$ depend on the maximum sequence length $n$.
In industrial settings with $n$ up to $100\mathrm{K}$ and highly variable user histories, this leads to large and inflexible parameterization.

\item \textbf{Handling variable-length sequences.}
User histories exhibit significant variation in length.
Explicit projections tied to a fixed $n$ require padding or truncation, which introduces inefficiencies and potential artifacts.

\item \textbf{Lack of target-agnostic reuse.}
The projected representation is typically embedded within an attention pipeline and does not naturally yield a reusable \emph{user-only} state that can be materialized once and cached across targets and consecutive requests.

\item \textbf{Mismatch to our system objective.}
In our setting, compression is not only a modeling device but also a systems primitive: the compressed representation must be fixed-size, target-agnostic, and cheap to reuse across both training-side MRLB and serving-side cache hits.
Explicit projection methods do not naturally provide this property.
\end{itemize}

These limitations motivate the need for a \emph{length-agnostic}, \emph{data-dependent}, and \emph{cacheable} compression mechanism.

\subsection{Sketch Attention as an Implicit Length Projection}

Sketch Attention (SA) constructs a compact representation of the history in the form
\begin{equation}
\widetilde{X} = A X,\qquad A\in\mathbb{R}^{k\times n},
\end{equation}
where $X\in\mathbb{R}^{n\times d}$ is the history embedding matrix and $A$ is a data-dependent assignment/projection matrix computed from $X$ and a set of learnable prototypes.

Specifically, SA defines
\begin{equation}
A = \softmax_{\text{proto}}\!\left(\frac{(PW_Q)(XW_K)^\top}{\sqrt{d}}\right),
\end{equation}
where $P\in\mathbb{R}^{k\times d}$ is a learnable prototype matrix.
The normalization is applied over the prototype dimension for each token, ensuring that each token distributes unit mass across prototypes.

This yields an assignment/projection matrix $A$ with the following properties:

\begin{itemize}
\item \textbf{Implicit and data-dependent.}
Unlike explicit $E,F$, the assignment/projection matrix $A$ is constructed dynamically from the input $X$, allowing it to adapt to different users and sequences.

\item \textbf{Length-agnostic parameterization.}
The learnable parameters consist of $P$, $W_Q$, and $W_K$, with total size $O(kd + d^2)$, independent of $n$.

\item \textbf{Fixed-size output.}
Regardless of the original sequence length, SA always produces a representation $\widetilde{X}\in\mathbb{R}^{k\times d}$.
This makes the downstream cost of sketch-side reasoning independent of $n$ once the sketch is available.

\item \textbf{End-to-end differentiability.}
The projection is fully differentiable and optimized jointly with downstream target-conditioned reasoning.

\item \textbf{Cacheability.}
Since $\widetilde{X}$ depends only on user-side signals, it can be computed once and reused across multiple targets and consecutive requests.
This property is central to our \emph{low-rank caching} view.
\end{itemize}

Therefore, SA can be viewed as realizing an \emph{implicit low-rank projection along the length dimension}: instead of explicitly learning $n\times k$ projection matrices, it constructs a compact, input-dependent map $A$ whose output size is fixed and whose parameters do not scale with the maximum sequence length.

\subsection{Relation and Difference to Linformer-Style Compression}

The connection to Linformer is primarily \emph{structural} rather than algorithmic.
Both views rely on the idea that long-sequence computation can be reduced through a compact representation along the length axis.
However, the two constructions differ in several important ways.

First, Linformer uses \emph{explicit} learned projections tied to the sequence length, whereas SA uses an \emph{implicit}, prototype-based, input-dependent projection.
Second, Linformer is motivated by approximating self-attention efficiently, whereas SA is designed as a target-agnostic \emph{compression front-end} for a downstream target-conditioned reasoning module.
Third, and most importantly for our setting, SA produces a reusable \emph{user-only} representation that can be cached and amortized across targets and requests, which is not a standard goal of low-rank attention approximations.

Thus, the low-rank interpretation should be understood as a representation-level analogy:
SA is not simply ``Linformer for recommendation,''
but a cacheable length-wise compression mechanism that plays a similar dimensionality-reduction role while serving a different system purpose.

\subsection{Discussion}

From this perspective, SequenceO1 follows a \emph{compress-then-reason} paradigm:
the ultra-long history is first mapped to a compact representation of length $k$, and target-conditioned reasoning is then performed on this fixed-size representation.

This view helps explain why the framework scales well to 100K histories.
The expensive dependence on the raw length $n$ is concentrated in the compression step, which is lightweight, target-agnostic, and reusable.
The heavyweight target-conditioned reasoning is then performed only on the compressed representation, whose size is fixed.
In this sense, the low-rank perspective is not merely a modeling analogy; it also provides the systems intuition behind SequenceO1:
by materializing and caching a compact length-wise representation, we turn ultra-long history modeling into a form of \emph{low-rank caching} suitable for industrial recommendation workloads.

\section{Algorithmic Summary of Sketch Attention}
\label{app:sa-algorithm}

Given history embeddings $X\in\mathbb{R}^{n\times d}$ and initial prototypes $P^{(0)}\in\mathbb{R}^{k\times d}$, one SA block proceeds as follows:
\begin{enumerate}
\item Compute prototype--token affinities:
\[
S=\frac{(P^{(0)}W_Q)(XW_K)^\top}{\sqrt{d}}.
\]
\item Normalize over prototypes for each token:
\[
A_{:,i}=\softmax_{\mathrm{proto}}(S_{:,i}),\qquad \sum_{j=1}^{k}A_{j,i}=1.
\]
\item Aggregate history tokens into sketch tokens:
\[
\widetilde{X}=AX.
\]
\item If stacked SA is used, apply residual connection, layer normalization, and FFN refinement, then repeat the allocation--aggregation step using the refined sketch state.
\end{enumerate}
The output is a fixed-size sketch $\widetilde{X}\in\mathbb{R}^{k\times d}$.

\section{Why Prototype-Wise Normalization}
\label{app:proto-norm}

This section explains why Sketch Attention (SA) normalizes over the \emph{prototype} dimension rather than the token dimension.
The key issue is that our compression module must be \emph{target-agnostic}: it is computed before seeing the candidate item, cached as a user-only representation, and then reused across many downstream targets and requests.
In this setting, the normalization choice determines whether the compressed representation tends to \emph{cover} the whole history or to \emph{select} only a small subset of tokens.

A natural alternative to SA is to normalize over the token dimension, yielding weights $p(i\mid j)$ for each prototype.
Under this scheme, each prototype attends to a subset of tokens and effectively acts as a selector over the sequence.
While this behavior can be useful when the downstream task already specifies what to focus on, it is less suitable for our setting because the compression must be computed \emph{before} target conditioning.

The problem is that token-wise normalization encourages \emph{competition among tokens for each prototype}, but does not enforce any corresponding coverage constraint over the full history.
As a result, multiple prototypes may collapse onto similar high-salience regions, repeatedly selecting the same small subset of tokens while ignoring others.
In a target-agnostic compression setting, this can lead to poor coverage: large portions of the ultra-long history may be underrepresented or discarded before the model has a chance to decide which parts will matter for a particular target.

In contrast, prototype-wise normalization produces weights $p(j\mid i)$ and enforces a per-token conservation constraint
\[
\sum_{j=1}^{k} A_{j,i}=1.
\]
This means that each token distributes its mass across prototypes, rather than each prototype independently selecting tokens.
Consequently, every token contributes mass to the compressed representation, and the prototypes collectively aggregate information from the entire history.

This distinction is important for SequenceO1.
Because the sketch must support diverse downstream targets and be reused across requests, it should preserve broad user-history coverage rather than prematurely commit to a small selected subset.
Prototype-wise normalization encourages this behavior: it turns the sketching step into a token-to-prototype \emph{allocation} mechanism, which is better aligned with reusable, user-only compression.

From this perspective, token-wise normalization is more naturally suited to \emph{target-aware selection}, whereas prototype-wise normalization is better suited to \emph{target-agnostic summarization}.
Our design choice reflects the role of SA in the overall framework: SA is not the final reasoning module, but a reusable compression layer whose output will later be consumed by target-conditioned STCA.
By preserving coverage early and deferring selective matching to the reasoning stage, SequenceO1 better balances information retention and computational efficiency.

\subsection{Assignment Patterns and Downstream Interfaces}
\label{app:sketch-discriminability}

The analysis in \S\ref{sec:exp-sa-vs-vanilla} uses the maximum normalized weight associated with each token as a descriptive diagnostic of the allocation pattern induced by each normalization rule.
Because prototype-wise and token-wise normalization operate over different axes, the resulting maxima are not calibrated confidence scores and are not interpreted as directly comparable measures of prototype specialization, overlap, or retrieval quality.
Within prototype-wise normalization, a larger maximum corresponds to a more concentrated token-to-prototype allocation, but it does not by itself guarantee diverse coverage or reduced prototype overlap.

This property is useful because the SA sketch is not the final reasoning output.
It is a reusable intermediate representation that will later be consumed by target-conditioned STCA.
These allocation patterns may provide an interface for future extensions, such as retrieving relevant behaviors through selected sketch tokens or applying localized refinement to behavior groups associated with a subset of prototypes.
These extensions are not evaluated in the current work; the task-level evidence for prototype-wise normalization comes from the UAUC improvements reported in Table~\ref{tab:uauc-sa-vs-vanilla}.

\section{Implementation Details of Action-Side Fusion}
\label{app:action-fusion}

Each historical behavior contains both item-side information and action-side information.
The action-side SwiGLU fusion used in Table~\ref{tab:sa-arch-ablation} denotes a lightweight gated fusion before sketch construction.
Given an item embedding $\mathbf{e}_i$ and an action-type embedding $\mathbf{a}_i$, the fused behavior embedding can be written abstractly as
\[
\mathbf{x}_i = \mathbf{e}_i + \mathrm{FFN}_{\mathrm{SwiGLU}}\!\left([\mathbf{e}_i;\mathbf{a}_i]\right),
\]
where the SwiGLU FFN maps the concatenated features back to the dimension of $\mathbf{e}_i$ so that the residual addition is well-defined; the exact production implementation may include additional feature transformations.
This fusion allows the same historical item to contribute differently under different feedback actions such as finish, skip, like, or follow.
The ablation ``w/o action-side SwiGLU fusion'' removes this action-aware gated fusion before SA.

\section{Iterative Refinement View}
\label{app:em}

This section provides an intuitive interpretation of stacked Sketch Attention (SA) as an iterative \emph{allocation--aggregation} procedure.
The goal is not to claim that SA exactly implements an optimization algorithm such as EM, but rather to highlight a useful analogy: each refinement step alternates between assigning history tokens to prototypes and updating the prototype-side representation by aggregating token information.

Consider one refinement step with current prototype representations $P$ (or, more generally, the current sketch state $\widetilde{X}^{(\ell)}$).
Using the current prototypes as queries, SA computes soft token-to-prototype assignments
\[
A_{j,i} = p(j\mid i),
\]
where each token distributes its mass across prototypes.
These assignments determine how the information in the ultra-long history is allocated to the fixed set of sketch slots.

Given the assignments, the updated sketch is obtained by aggregating token embeddings:
\[
\widetilde{X} = A X.
\]
Thus, one SA step can be viewed as first deciding \emph{how} each token contributes to different prototypes, and then recomputing each prototype representation as a weighted summary of the full history.

With stacked SA, this process is repeated multiple times.
After one round of allocation and aggregation, the prototype-side representation becomes more informed by the input history.
Using this refined representation to compute the next round of assignments allows the model to adjust how tokens are grouped and summarized.
In this sense, deeper SA layers act as refinement steps that progressively improve the sketch, rather than as entirely independent transformations.

This view is reminiscent of an EM-like procedure:
the soft assignments play a role analogous to a responsibility update, while the aggregation step plays a role analogous to prototype re-estimation.
The analogy is only conceptual---SA is trained end-to-end with gradient descent, includes residual connections and feed-forward layers, and is not derived from a probabilistic latent-variable objective.
Still, the EM perspective is useful because it explains why stacked SA can improve sketch quality: later layers refine the token-to-prototype allocation using prototypes that already encode information from earlier aggregation steps.

This interpretation also helps motivate why only a small number of refinement steps is sufficient in practice.
User behavior sequences are highly redundant, and the purpose of SA is not to recover fine-grained token-level structure, but to construct a compact user-level summary that preserves the most salient long-range signals.
As a result, substantial gains can often be obtained after only one or two rounds of refinement, after which additional iterations yield diminishing returns.
This matches our empirical finding that a small stacked depth ($N_{\mathrm{sa}}=2$) is already effective for 100K sequence modeling.

\section{Error Decomposition}
\label{app:error}

This section provides a simple conceptual view of the approximation error introduced by the \emph{compress-then-reason} design in SequenceO1.
Our goal is not to derive a tight theoretical bound, but to clarify that the overall error can be separated into two sources:
(i) information loss caused by compressing the ultra-long history into a fixed-size sketch, and
(ii) modeling error incurred when performing target-conditioned reasoning on the compressed representation.

Let $F_{\mathrm{full}}(x_t, X)$ denote an ideal target-conditioned predictor that operates directly on the full history $X$.
Let $F_{\mathrm{comp}}^\star(x_t,C(X))$ denote the best predictor within a chosen hypothesis class that has access only to the compressed representation $C(X)$.
Our method applies a downstream reasoning module $G(\cdot,\cdot)$:
\[
\hat F(x_t, X) = G(x_t, C(X)).
\]
In our setting, $C(X)$ corresponds to the Sketch Attention (SA) sketch, while $G$ corresponds to the subsequent target-conditioned reasoning and fusion modules.

To expose the approximation structure, add and subtract $F_{\mathrm{comp}}^\star(x_t,C(X))$:
\[
F_{\mathrm{full}}(x_t,X)-\hat F(x_t,X)
=
\big(F_{\mathrm{full}}(x_t,X)-F_{\mathrm{comp}}^\star(x_t,C(X))\big)
+
\big(F_{\mathrm{comp}}^\star(x_t,C(X))-\hat F(x_t,X)\big).
\]
Applying the triangle inequality gives
\[
\|F_{\mathrm{full}}(x_t,X)-\hat F(x_t,X)\|
\le
\underbrace{\|F_{\mathrm{full}}(x_t,X)-F_{\mathrm{comp}}^\star(x_t,C(X))\|}_{\text{compression error}}
+
\underbrace{\|F_{\mathrm{comp}}^\star(x_t,C(X))-\hat F(x_t,X)\|}_{\text{reasoning error}}.
\]

The first term measures how much predictive information is lost when replacing the full history $X$ with its compressed representation $C(X)$.
This term is controlled by the quality of the sketch: if the ultra-long history is sufficiently compressible along the length dimension, and the sketch preserves the salient user signals relevant to downstream ranking, then this term can remain small even when $C(X)$ has fixed size.

The second term measures how well the downstream model $G$ can exploit the compressed representation once it is available.
Even if the sketch preserves most useful information, insufficient downstream capacity or suboptimal target-conditioned interaction can still lead to prediction error.
In SequenceO1, this term is addressed by applying STCA over both the recent 10K suffix and the ultra-long sketch, followed by lightweight fusion.

This decomposition helps explain the design principle behind SequenceO1.
SA is used to control the \emph{compression error} by constructing a compact but expressive user-only summary of the 100K history, while the subsequent STCA-based reasoning stack controls the \emph{reasoning error} by allocating model capacity to target-aware interaction rather than to repeatedly scanning the raw ultra-long sequence.
From this perspective, the effectiveness of the overall framework depends on balancing these two terms: the sketch must be compact enough for efficiency and reuse, yet informative enough that downstream reasoning over the sketch remains highly predictive.

\section{Complexity Derivation (FLOPs)}
\label{app:complexity}

This appendix derives the FLOP formulas used in our complexity analysis under the notation of the main paper; $L_s$ denotes the sketch length $k$ in the main text.
\textbf{SA is single-head} and operates on width $d=128$.
\textbf{STCA is multi-head} with $h=16$ heads and head dimension $d_h=64$, hence the STCA model width is
\begin{equation}
D \triangleq h\cdot d_h = 1024.
\end{equation}
All FLOPs are reported under the same GEMM-dominant counting convention.

\subsection{Counting convention}
\label{app:flop_convention}

We report complexity in terms of FLOPs dominated by matrix multiplications (GEMMs).
For a matrix product $U\in\mathbb{R}^{m\times r}$ and $V\in\mathbb{R}^{r\times p}$,
\begin{equation}
\mathrm{FLOPs}(UV)\triangleq 2mrp,
\label{eq:flops_gemm}
\end{equation}
counting one multiply and one add per inner-dimension element.
We ignore lower-order elementwise operations such as softmax exponentials, LayerNorm, bias adds, masking, and activation function costs.

\paragraph{SwiGLU FFN FLOPs.}
For a SwiGLU FFN applied to $T$ tokens with model width $w$ and intermediate width $w_f$,
\begin{equation}
\mathrm{FFN}(U)=\big(\mathrm{swish}(UW_1)\odot (UW_g)\big)W_2,
\end{equation}
with $W_1,W_g\in\mathbb{R}^{w\times w_f}$ and $W_2\in\mathbb{R}^{w_f\times w}$, the GEMM FLOPs are
\begin{equation}
\mathrm{FLOPs}_{\mathrm{SwiGLU}}(T;w,w_f)=2Tww_f+2Tww_f+2Tw_fw=6Tww_f.
\label{eq:flops_swiglu}
\end{equation}
In our setting we use $w_f=2w$ (both for SA and STCA FFNs), hence
\begin{equation}
\mathrm{FLOPs}_{\mathrm{SwiGLU}}(T;w,2w)=12Tw^2.
\label{eq:flops_swiglu_2w}
\end{equation}

\subsection{STCA FLOPs (with single-query reordering)}
\label{app:stca_flops}

STCA uses stacked single-query target-to-history cross attention.
Let the history length be $L$ and the STCA model width be $D$.
We use $h$ heads with per-head dimension $d_h=D/h$, and denote the number of STCA layers by $N_{\mathrm{stca}}$.
In addition to attention and the query-side FFN, each STCA layer includes its own
\textbf{history-side pre-processing SwiGLU FFN} applied to all $L$ history tokens before that layer interacts with the target
(i.e., a target-independent history transform), with intermediate width $2D$ and output width $D$.

\paragraph{History-side pre-FFN (per layer).}
In each layer, the history-side SwiGLU is applied to $X\in\mathbb{R}^{L\times D}$:
\begin{equation}
\mathrm{FLOPs}_{\mathrm{hist\text{-}FFN}}(L;D)
=
\mathrm{FLOPs}_{\mathrm{SwiGLU}}(L;D,2D)
=
12LD^2.
\label{eq:stca_hist_ffn}
\end{equation}

\subsubsection{Single-query attention reordering and FLOPs analysis}
\label{app:attn-opt}

With exactly one query per layer, let $X\!\in\!\mathbb{R}^{L\times D}$, $q\!\in\!\mathbb{R}^{1\times D}$, and $d_h{=}D/h$.
The standard cross-attention form is
\begin{equation}
\label{eq:attn-standard}
\mathrm{Attn}(q,X)
\;=\;
\softmax\!\Big(\tfrac{(qW_Q)(XW_K)^\top}{\sqrt{d_h}}\Big)\cdot(XW_V),
\end{equation}
which projects all $L$ tokens twice and materializes the length-$L$ tensors $XW_K$ and $XW_V$.

\paragraph{Reordering.}
We can reorder the computation to remove the length-$L$ projections:
\[
u \,=\, (qW_Q)W_K^\top \in \mathbb{R}^{1\times D},\quad
\alpha \,=\, \softmax\!\Big(\tfrac{u\,X^\top}{\sqrt{d_h}}\Big)\in\mathbb{R}^{1\times L},
\]
and then compute
\[
o \,=\, (\alpha X)\,W_V \in \mathbb{R}^{1\times d_h},
\quad W_Q,W_K,W_V\!\in\!\mathbb{R}^{D\times d_h}.
\]
Equivalently,
\begin{equation*}
\mathrm{Attn}(q,X)
\;=\;
\Big(\softmax\!\big(\tfrac{((qW_Q)W_K^\top)X^\top}{\sqrt{d_h}}\big)\,X\Big)W_V
\;=\;(\alpha X)W_V .
\end{equation*}

\paragraph{FLOPs (GEMM-dominant).}
We count GEMMs and ignore softmax/LN/activation overhead.
Per head, the reordered path costs
\[
\underbrace{2Dd_h}_{qW_Q}
+\underbrace{2Dd_h}_{(qW_Q)W_K^\top}
+\underbrace{2LD}_{uX^\top}
+\underbrace{2LD}_{\alpha X}
+\underbrace{2Dd_h}_{(\alpha X)W_V},
\]
so across $h$ heads the attention FLOPs per layer are
\begin{equation}
\mathrm{FLOPs}_{\mathrm{attn\text{-}reorder}}(L;D,h)
=
\underbrace{4LDh}_{\text{length-dependent}}
+
\underbrace{6D^2}_{\text{head projections}}.
\label{eq:stca_attn_reorder}
\end{equation}
Compared with the na\"ive path that explicitly forms $(XW_K, XW_V)$ (length-dependent cost $\approx 4LD^2$),
the reordered path removes the $O(LD^2)$ projections and replaces them with $O(LDh)$ weighted reductions,
reducing the length-dependent FLOPs by a factor of approximately $d_h=D/h$.

\paragraph{No cross-request KV sharing under reordering.}
Importantly, reordering avoids materializing the $L{\times}d_h$ intermediates $(XW_K, XW_V)$, so there are \emph{no}
target-independent per-layer KV projections to cache/share across requests.
The remaining $L$-dependent terms ($uX^\top$ and $\alpha X$) depend on the query and thus are inherently per-target.

\paragraph{Other per-layer query-side transforms.}
Following our main accounting convention, we further include:
(i) an output projection $W_O\in\mathbb{R}^{D\times D}$ with FLOPs $2D^2$,
and (ii) a one-token query-side SwiGLU FFN (intermediate width $2D$) with FLOPs $12D^2$.

\paragraph{Per-layer STCA FLOPs (optimized).}
Combining Eq.~\eqref{eq:stca_attn_reorder} with the output projection and query FFN, the per-layer per-target FLOPs are
\begin{equation}
\mathrm{FLOPs}_{\mathrm{STCA\text{-}layer}}(L;D,h)
=
\mathrm{FLOPs}_{\mathrm{attn\text{-}reorder}}(L;D,h)
+
2D^2
+
12D^2
=
4LDh + 20D^2.
\label{eq:stca_layer_reorder_total}
\end{equation}

\paragraph{Total STCA FLOPs (per target).}
Including the independent history-side pre-FFN in every layer, the total per-target STCA FLOPs are
\begin{equation}
\mathrm{FLOPs}_{\mathrm{STCA}}(L;N_{\mathrm{stca}},D,h)
=
N_{\mathrm{stca}}\left(12LD^2 + 4LDh + 20D^2\right).
\label{eq:stca_total_reorder}
\end{equation}

\subsection{Sketch Attention (SA) FLOPs (single-head)}
\label{app:sa_flops}

SA compresses an ultra-long history $X\in\mathbb{R}^{L\times d}$ into a fixed-size sketch
$\widetilde{X}\in\mathbb{R}^{L_s\times d}$ (\S\ref{sec:method-sa}),
where $L_s$ denotes the number of prototypes (i.e., the sketch length).
We assume SA is \textbf{single-head} in our instantiation and uses width $d=128$.

\paragraph{One SA layer.}
Let prototypes be $P\in\mathbb{R}^{L_s\times d}$ and projections $W_Q,W_K\in\mathbb{R}^{d\times d}$:
\begin{equation}
Q=PW_Q\in\mathbb{R}^{L_s\times d},\qquad
K=XW_K\in\mathbb{R}^{L\times d}.
\end{equation}
The affinity matrix is $S=\frac{QK^\top}{\sqrt{d}}\in\mathbb{R}^{L_s\times L}$ and the assignment
$A=\softmax_{\text{proto}}(S)\in\mathbb{R}^{L_s\times L}$.
The sketch is
\begin{equation}
\widetilde{X}=AX \in \mathbb{R}^{L_s\times d}.
\label{eq:sa_sketch_agg}
\end{equation}

\paragraph{SA layer FLOPs (GEMM-dominant).}
We count:
\begin{itemize}
\item \textbf{Key projection:} $K=XW_K$:
\begin{equation}
\mathrm{FLOPs}_{K}^{\mathrm{SA}}(L;d)=2Ld^2.
\end{equation}
\item \textbf{Prototype projection:} $Q=PW_Q$:
\begin{equation}
\mathrm{FLOPs}_{Q}^{\mathrm{SA}}(L_s;d)=2L_s d^2.
\end{equation}
\item \textbf{Affinity GEMM:} $S=QK^\top$:
\begin{equation}
\mathrm{FLOPs}_{\mathrm{score}}^{\mathrm{SA}}(L,L_s;d)=2L_sLd.
\end{equation}
\item \textbf{Aggregation GEMM:} $\widetilde{X}=AX$:
\begin{equation}
\mathrm{FLOPs}_{\mathrm{agg}}^{\mathrm{SA}}(L,L_s;d)=2L_sLd.
\end{equation}
\item \textbf{Sketch FFN:} SwiGLU applied to $L_s$ sketch tokens with intermediate width $2d$:
\begin{equation}
\mathrm{FLOPs}_{\mathrm{FFN}}^{\mathrm{SA}}(L_s;d)=\mathrm{FLOPs}_{\mathrm{SwiGLU}}(L_s;d,2d)=12L_s d^2.
\end{equation}
\end{itemize}

Summing these yields the per-layer SA FLOPs:
\begin{equation}
\mathrm{FLOPs}_{\mathrm{SA\text{-}layer}}(L,L_s;d)
=
2Ld^2 + 2L_s d^2 + 4L_sLd + 12L_s d^2
=
2Ld^2 + 14L_s d^2 + 4L_sLd.
\label{eq:sa_layer_total}
\end{equation}

\paragraph{Stacked SA.}
With $N_{\mathrm{sa}}$ stacked SA layers (\S\ref{sec:method-stack}), the total SA cost is
\begin{equation}
\mathrm{FLOPs}_{\mathrm{SA}}(L;N_{\mathrm{sa}},L_s,d)=N_{\mathrm{sa}}\cdot \mathrm{FLOPs}_{\mathrm{SA\text{-}layer}}(L,L_s;d).
\label{eq:sa_total}
\end{equation}

\subsection{Single-branch SA+STCA on the sketch: cache miss/hit and expected cost}
\label{app:sa_stca_cache}

We analyze the \emph{ultra-long single branch} used in Fig.~\ref{fig:complexity}, consisting of SA sketching over $L$ tokens followed by an $N_{\mathrm{stca}}$-layer STCA over the fixed-size sketch of length $L_s$.
Common costs outside this branch---including the recent-10K STCA branch, dense-ranker computation, input embedding and action-feature preprocessing, and final two-branch fusion---are omitted from both alternatives.
This FLOP analysis is conservative with respect to system savings: it does not count the raw-length feature materialization, storage, and communication that are also bypassed when a cached sketch is available.
Since SA uses width $d$ while STCA uses width $D$, we apply a lightweight linear adapter to map the sketch width.
The target-side projection $\phi_t(\mathbf{x}_t)$ is candidate-side and independent of the history length, so it is omitted from the sequence-branch FLOP accounting. Here $W_A$ denotes the width adapter and is distinct from the SA assignment matrix $A$:
\begin{equation}
\widehat{X} = \widetilde{X}W_A,\qquad W_A\in\mathbb{R}^{d\times D},\qquad \widehat{X}\in\mathbb{R}^{L_s\times D}.
\end{equation}
Only the $d$-width SA sketch is stored in the persistent cache; the adapter output is \textbf{not persistently cached across groups}.
Its cost is therefore incurred on both cache hits and misses, although the target-independent adapted sketch is materialized once and reused within an MRLB group.

\paragraph{Adapter FLOPs.}
\begin{equation}
\mathrm{FLOPs}_{\mathrm{adapt}}(L_s;d,D)=2L_s d D.
\label{eq:adapt_flops}
\end{equation}

\paragraph{Hit and miss FLOPs.}
On a cache miss, we compute SA, adapt the sketch, and then run sketch-STCA; on a cache hit, we reuse the cached sketch and run the same adapt+STCA:
\begin{align}
\mathrm{FLOPs}_{\mathrm{miss}}(L)
&=
\mathrm{FLOPs}_{\mathrm{SA}}(L;N_{\mathrm{sa}},L_s,d)
+\mathrm{FLOPs}_{\mathrm{adapt}}(L_s;d,D)
+\mathrm{FLOPs}_{\mathrm{STCA}}(L_s;N_{\mathrm{stca}},D,h),
\label{eq:cache_miss}\\
\mathrm{FLOPs}_{\mathrm{hit}}
&=
\mathrm{FLOPs}_{\mathrm{adapt}}(L_s;d,D)
+\mathrm{FLOPs}_{\mathrm{STCA}}(L_s;N_{\mathrm{stca}},D,h).
\label{eq:cache_hit}
\end{align}

\paragraph{Expected FLOPs under hit rate.}
Let $p_{\mathrm{hit}}\in[0,1]$ be the cache hit probability. The expected FLOPs are
\begin{equation}
\mathbb{E}[\mathrm{FLOPs}](L)
=
p_{\mathrm{hit}}\cdot \mathrm{FLOPs}_{\mathrm{hit}}
+(1-p_{\mathrm{hit}})\cdot \mathrm{FLOPs}_{\mathrm{miss}}(L)
=
\mathrm{FLOPs}_{\mathrm{hit}} + (1-p_{\mathrm{hit}})\cdot \mathrm{FLOPs}_{\mathrm{SA}}(L;N_{\mathrm{sa}},L_s,d).
\label{eq:expected_flops}
\end{equation}

\subsection{Hyperparameter instantiation for analysis}
\label{app:hparams}

We instantiate the formulas above with the following settings:
\begin{itemize}
\item \textbf{SA width:} $d=128$ (single-head SA and its SwiGLU uses intermediate width $2d$).
\item \textbf{STCA heads:} $h=16$, $d_h=64$, hence $D=hd_h=1024$ (STCA SwiGLU uses intermediate width $2D$).
\item \textbf{SA:} $N_{\mathrm{sa}}=2$ stacked SA layers, single-head; number of prototypes / sketch length $L_s=1024$.
\item \textbf{STCA:} $N_{\mathrm{stca}}=4$ layers.
\item \textbf{Cache hit rate:} for training-side analysis we use $p_{\mathrm{hit}}^{\mathrm{train}}=0.5$; for inference/serving-side analysis we use $p_{\mathrm{hit}}^{\mathrm{infer}}=0.6$.
\item \textbf{Reuse ratio:} for training-side MRLB amortization we use $R_{\mathrm{train}}=40$; for inference/serving-side amortization across consecutive requests we use $R_{\mathrm{infer}}=300$.
\item \textbf{Sequence lengths:} $L \in \{1\mathrm{K},2\mathrm{K},10\mathrm{K},20\mathrm{K},50\mathrm{K},100\mathrm{K}\}$ in our plots.
\end{itemize}

\paragraph{Closed forms.}
With the above settings, we obtain:
\begin{align}
\mathrm{FLOPs}_{\mathrm{STCA}}(L;N_{\mathrm{stca}},D,h)
&=
N_{\mathrm{stca}}\left(12LD^2 + 4LDh + 20D^2\right),
\label{eq:stca_closed}\\
\mathrm{FLOPs}_{\mathrm{SA}}(L;N_{\mathrm{sa}},L_s,d)
&=
N_{\mathrm{sa}}\left(2Ld^2 + 14L_s d^2 + 4L_sLd\right),
\label{eq:sa_closed}\\
\mathrm{FLOPs}_{\mathrm{hit}}
&=
\mathrm{FLOPs}_{\mathrm{adapt}}(L_s;d,D)+\mathrm{FLOPs}_{\mathrm{STCA}}(L_s;N_{\mathrm{stca}},D,h),
\qquad
\mathrm{FLOPs}_{\mathrm{miss}}(L)=\mathrm{FLOPs}_{\mathrm{SA}}(L;N_{\mathrm{sa}},L_s,d)+\mathrm{FLOPs}_{\mathrm{hit}},
\label{eq:hit_miss_closed}\\
\mathbb{E}[\mathrm{FLOPs}](L;p_{\mathrm{hit}})
&=
\mathrm{FLOPs}_{\mathrm{hit}} + (1-p_{\mathrm{hit}})\cdot \mathrm{FLOPs}_{\mathrm{SA}}(L;N_{\mathrm{sa}},L_s,d).
\label{eq:expected_closed}
\end{align}

\paragraph{Intuitive comparison and dominant terms.}
For \textbf{STCA-only} on length $L$, the dominant cost consists of the per-layer history-side pre-FFN term $12LD^2$ and the per-layer query--history interaction term $4LDh$ under single-query reordering (Eq.~\eqref{eq:stca_closed}). Thus, STCA-only scales linearly in both $L$ and $N_{\mathrm{stca}}$, with the large model width $D$ appearing in both the history-side transform and the per-layer query-side transforms.

For \textbf{SA}, the dominant term is the prototype--token interaction $4L_sLd$ in each SA layer
(from $QK^\top$ and $AX$), yielding $O(N_{\mathrm{sa}}L_sLd)$ overall (Eq.~\eqref{eq:sa_closed}).
Therefore, SA also scales linearly in $L$, with slope controlled by the fixed sketch length $L_s$ and the smaller SA width $d$.

\paragraph{Why both are $O(L)$ but behave very differently in practice.}
Although both SA and STCA have linear dependence on the input length $L$, they allocate modeling capacity very differently.
SA is designed as a \emph{lightweight} length-wise compressor: it operates at a much smaller width $d$ and summarizes the $L$ history tokens into
a fixed-size sketch of length $L_s$, which is consistent with the empirically observed \emph{compressibility} of long behavioral sequences
along the length dimension. In contrast, STCA is the \emph{heavyweight} target-conditioned reasoning module: it uses a substantially larger width $D$
(and multi-head structure) to model fine-grained interactions between the target and the (raw or compressed) history.
This design intentionally places expensive representational power in the target--history interaction stage, where precise matching is required,
while keeping the long-history compression stage efficient.

For \textbf{SA+STCA}, sketch-STCA runs on the fixed length $L_s$.
On a \textbf{cache hit}, the cost is independent of $L$ and consists of the (uncached) adapter plus STCA on length $L_s$
(Eq.~\eqref{eq:cache_hit}); on a \textbf{cache miss}, the additional $L$-dependent cost comes entirely from SA (Eq.~\eqref{eq:cache_miss}).
With hit rate $p_{\mathrm{hit}}$, the expected compute equals the hit cost plus a reduced linear-in-$L$ component weighted by $(1-p_{\mathrm{hit}})$ (Eq.~\eqref{eq:expected_flops}).

\subsection{MRLB Amortization and Reuse Ratio}
\label{app:mrlb_amort}

We further analyze the per-target FLOPs under Multi-Request Level Batching (MRLB), where $R$ denotes the number of target instances across grouped requests that share the same user-side computation.
Within an MRLB group, target-independent (user-only) computations can be executed once and reused across these $R$ target instances.
Thus, if a computation can be decomposed as
\begin{equation}
\mathrm{FLOPs}_{\mathrm{total}} = \mathrm{FLOPs}_{\mathrm{shared}} + \mathrm{FLOPs}_{\mathrm{per\text{-}target}},
\end{equation}
then the \emph{amortized} per-target FLOPs under MRLB are
\begin{equation}
\mathrm{FLOPs}_{\mathrm{MRLB}}(R) = \frac{\mathrm{FLOPs}_{\mathrm{shared}}}{R} + \mathrm{FLOPs}_{\mathrm{per\text{-}target}}.
\label{eq:mrlb_rule}
\end{equation}

\paragraph{STCA-only under MRLB (single-query reordering).}
For STCA with model width $D$, $h$ heads, and $N_{\mathrm{stca}}$ layers, we adopt the single-query attention reordering
described in \S\ref{app:attn-opt}.
Under this optimized implementation, the $L\times d_h$ intermediates $(XW_K, XW_V)$ are \emph{not} materialized, i.e.,
there are \emph{no} explicit history-side per-layer $K/V$ projections that can be cached or reused across requests.
Consequently, within MRLB the only strictly target-independent components we amortize are the \textbf{per-layer history-side pre-FFNs}.

Concretely, the \emph{shared} FLOPs are
\begin{equation}
\mathrm{FLOPs}_{\mathrm{STCA,shared}}(L)
=
12N_{\mathrm{stca}}LD^2,
\label{eq:stca_shared_mrlb}
\end{equation}
and the remaining \emph{per-target} FLOPs are the per-layer reordered attention plus query-side transforms.
Using Eq.~\eqref{eq:stca_attn_reorder} and Eq.~\eqref{eq:stca_layer_reorder_total}, we have
\begin{equation}
\mathrm{FLOPs}_{\mathrm{STCA,per\text{-}target}}(L)
=
N_{\mathrm{stca}}\left(4LDh + 20D^2\right).
\label{eq:stca_pertarget_mrlb}
\end{equation}
Therefore, the amortized per-target STCA-only FLOPs under MRLB are
\begin{equation}
\mathrm{FLOPs}^{\mathrm{MRLB}}_{\mathrm{STCA}}(L;N_{\mathrm{stca}},D,h,R)
=
\frac{12N_{\mathrm{stca}}LD^2}{R}
+
N_{\mathrm{stca}}\left(4LDh + 20D^2\right).
\label{eq:stca_mrlb_total}
\end{equation}

\paragraph{SA+Adapter+STCA on the sketch under MRLB (hit/miss).}
We analyze the ultra-long single branch consisting of SA sketching over length $L$ (SA width $d$), an adapter $d\!\to\!D$,
and sketch-STCA over length $L_s$ (STCA width $D$).
Under MRLB, we assume:
(i) \textbf{all SA computation is reusable} within the group (user-only);
(ii) the adapter output is target-independent and thus reusable within the group; and
(iii) sketch-STCA uses the same single-query reordering as STCA, so there are \emph{no} per-layer sketch-side $K/V$ projections
to reuse across requests; only the per-layer sketch-side pre-FFNs are reusable.

Let the adapter FLOPs be Eq.~\eqref{eq:adapt_flops}.
For sketch-STCA at length $L_s$, the reusable and per-target parts follow Eq.~\eqref{eq:stca_shared_mrlb}--\eqref{eq:stca_pertarget_mrlb}
by substituting $L\leftarrow L_s$:
\begin{align}
\mathrm{FLOPs}_{\mathrm{sk\text{-}STCA,shared}}
&=
12N_{\mathrm{stca}}L_sD^2,
\label{eq:skstca_shared_mrlb}\\
\mathrm{FLOPs}_{\mathrm{sk\text{-}STCA,per\text{-}target}}
&=
N_{\mathrm{stca}}\left(4L_sDh + 20D^2\right).
\label{eq:skstca_pertarget_mrlb}
\end{align}

On a \textbf{cache hit}, SA is skipped but we still perform adapter + sketch-STCA:
\begin{equation}
\mathrm{FLOPs}^{\mathrm{MRLB}}_{\mathrm{hit}}(R)
=
\frac{\mathrm{FLOPs}_{\mathrm{adapt}}(L_s;d,D)+\mathrm{FLOPs}_{\mathrm{sk\text{-}STCA,shared}}}{R}
+
\mathrm{FLOPs}_{\mathrm{sk\text{-}STCA,per\text{-}target}}.
\label{eq:hit_mrlb}
\end{equation}

On a \textbf{cache miss}, SA is computed once and reused across the $R$ targets:
\begin{equation}
\mathrm{FLOPs}^{\mathrm{MRLB}}_{\mathrm{miss}}(L;R)
=
\frac{\mathrm{FLOPs}_{\mathrm{SA}}(L;N_{\mathrm{sa}},L_s,d)+\mathrm{FLOPs}_{\mathrm{adapt}}(L_s;d,D)+\mathrm{FLOPs}_{\mathrm{sk\text{-}STCA,shared}}}{R}
+
\mathrm{FLOPs}_{\mathrm{sk\text{-}STCA,per\text{-}target}}.
\label{eq:miss_mrlb}
\end{equation}

\paragraph{Expected FLOPs with cache hit rate under MRLB.}
Here $p_{\mathrm{hit}}$ denotes the probability that the cached user sketch is available for an MRLB group; the corresponding hit or miss is shared by the $R$ target instances in that group.
The expected per-target FLOPs become
\begin{equation}
\mathbb{E}\!\left[\mathrm{FLOPs}^{\mathrm{MRLB}}\right](L;R,p_{\mathrm{hit}})
=
\mathrm{FLOPs}_{\mathrm{sk\text{-}STCA,per\text{-}target}}
+
\frac{\mathrm{FLOPs}_{\mathrm{adapt}}(L_s;d,D)+\mathrm{FLOPs}_{\mathrm{sk\text{-}STCA,shared}} + (1-p_{\mathrm{hit}})\cdot \mathrm{FLOPs}_{\mathrm{SA}}(L;N_{\mathrm{sa}},L_s,d)}{R}.
\label{eq:expected_mrlb}
\end{equation}

\subsection{Cache-Hit Sensitivity and Miss Cost}
\label{app:cache_sensitivity}

Equations~\eqref{eq:expected_flops} and~\eqref{eq:expected_mrlb} make the dependence on the cache hit rate explicit.
Without MRLB, a cache miss adds exactly the SA sketching cost:
\begin{equation}
\mathrm{FLOPs}_{\mathrm{miss}}(L)-\mathrm{FLOPs}_{\mathrm{hit}}
=
\mathrm{FLOPs}_{\mathrm{SA}}(L;N_{\mathrm{sa}},L_s,d),
\end{equation}
and the expected cost changes linearly with $1-p_{\mathrm{hit}}$:
\begin{equation}
\mathbb{E}[\mathrm{FLOPs}](L;p_{\mathrm{hit}})
=
\mathrm{FLOPs}_{\mathrm{hit}}
+
(1-p_{\mathrm{hit}})\mathrm{FLOPs}_{\mathrm{SA}}(L;N_{\mathrm{sa}},L_s,d).
\end{equation}

Under MRLB, the miss penalty is further amortized by the reuse ratio $R$:
\begin{equation}
\mathrm{FLOPs}^{\mathrm{MRLB}}_{\mathrm{miss}}(L;R)
-
\mathrm{FLOPs}^{\mathrm{MRLB}}_{\mathrm{hit}}(R)
=
\frac{\mathrm{FLOPs}_{\mathrm{SA}}(L;N_{\mathrm{sa}},L_s,d)}{R}.
\end{equation}
Therefore, decreasing the hit rate from $p_1$ to $p_2$ increases the amortized per-target cost by
\begin{equation}
\Delta \mathrm{FLOPs}^{\mathrm{MRLB}}
=
\frac{(p_1-p_2)\mathrm{FLOPs}_{\mathrm{SA}}(L;N_{\mathrm{sa}},L_s,d)}{R}.
\end{equation}

The all-miss case corresponds to $p_{\mathrm{hit}}=0$:
\begin{equation}
\mathbb{E}\!\left[\mathrm{FLOPs}^{\mathrm{MRLB}}\right](L;R,0)
=
\mathrm{FLOPs}_{\mathrm{sk\text{-}STCA,per\text{-}target}}
+
\frac{\mathrm{FLOPs}_{\mathrm{adapt}}(L_s;d,D)+\mathrm{FLOPs}_{\mathrm{sk\text{-}STCA,shared}}+\mathrm{FLOPs}_{\mathrm{SA}}(L;N_{\mathrm{sa}},L_s,d)}{R}.
\end{equation}

This analysis clarifies the role of caching.
Cache hits remove the raw-length SA sketching cost and, at the system level, bypass raw-history feature materialization, storage, and communication; cache misses reintroduce only the lightweight user-only SA computation.
In both cases, the heavyweight target-conditioned STCA branch operates on the fixed sketch length $L_s$ rather than the raw length $L$.
Thus, lower hit rates reduce the amortization benefit smoothly, but do not turn SequenceO1 back into target-conditioned reasoning over the raw 100K history.

\paragraph{Discussion.}
With single-query reordering, STCA removes the explicit $O(LD^2)$ history-side $K/V$ projections and thus cannot amortize them via MRLB;
only the target-independent per-layer pre-FFN term $12N_{\mathrm{stca}}LD^2$ is reduced by a factor of $R$.
The remaining per-target attention interaction scales as $O(N_{\mathrm{stca}}LDh)$ and is query-dependent.
In contrast, SA+STCA can amortize the \emph{entire} SA sketching stage by $R$, and SA’s dominant term scales as $O(N_{\mathrm{sa}}L_sLd)$ with a
much smaller width $d$.
This further highlights the \emph{compress-then-reason} design: long-history processing is pushed into a lightweight, highly reusable stage (SA),
while the heavyweight capacity (large $D$) is concentrated in target-conditioned interaction (STCA) where finer modeling is required.

\paragraph{FLOP-ratio comparison at 100K: without MRLB, training, and inference.}
We quantify the relative compute gap between the direct-STCA and SequenceO1 ultra-long branches at the 100K regime using FLOP
ratios under the same GEMM-only counting convention (lower-order ops ignored).
Let the STCA history length be $L=100\mathrm{K}$, the sketch length be $L_s=1024$, the STCA width be $D=hd_h=1024$
with $h=16$ heads and $N_{\mathrm{stca}}=4$ layers, and the SA width be $d=128$ with $N_{\mathrm{sa}}=2$ layers.
We use a training-side cache TTL of 3h with hit rate $p_{\mathrm{hit}}^{\mathrm{train}}=0.5$, an inference/serving-side cache TTL of 1h with hit rate $p_{\mathrm{hit}}^{\mathrm{infer}}=0.6$, training-side reuse ratio $R_{\mathrm{train}}=40$, and inference/serving-side reuse ratio $R_{\mathrm{infer}}=300$.

\paragraph{Without MRLB (reuse ratio 1).}
Under single-query reordering (\S\ref{app:attn-opt}), STCA-only FLOPs at length $L$ are
\begin{equation}
\mathrm{FLOPs}_{\mathrm{STCA}}(L)=N_{\mathrm{stca}}(12LD^2+4LDh+20D^2).
\end{equation}
SequenceO1 uses SA sketching (Eq.~\eqref{eq:sa_closed}), an uncached adapter (Eq.~\eqref{eq:adapt_flops}), and sketch-STCA
at length $L_s$ using the same reordering:
\begin{align}
\mathrm{FLOPs}_{\mathrm{SeqO1,hit}}
&=
\mathrm{FLOPs}_{\mathrm{adapt}}(L_s;d,D)
+
\mathrm{FLOPs}_{\mathrm{STCA}}(L_s),\\
\mathrm{FLOPs}_{\mathrm{SeqO1,miss}}(L)
&=
\mathrm{FLOPs}_{\mathrm{SA}}(L;N_{\mathrm{sa}},L_s,d)
+
\mathrm{FLOPs}_{\mathrm{SeqO1,hit}},\\
\mathbb{E}[\mathrm{FLOPs}]_{\mathrm{SeqO1}}(L;p_{\mathrm{hit}})
&=
\mathrm{FLOPs}_{\mathrm{SeqO1,hit}}
+
(1-p_{\mathrm{hit}})\cdot \mathrm{FLOPs}_{\mathrm{SA}}(L;N_{\mathrm{sa}},L_s,d).
\end{align}
At $L=100\mathrm{K}$, plugging in the training-side cache hit rate $p_{\mathrm{hit}}^{\mathrm{train}}=0.5$ yields
\[
\mathrm{STCA}\approx 5059.46\ \mathrm{GFLOPs/target},\quad
\mathrm{SeqO1}_{\mathrm{exp,train\text{-}hit}}\approx 108.10\ \mathrm{GFLOPs/target},
\]
so the expected FLOP ratio without MRLB under the training-side hit-rate assumption is
\begin{equation}
\mathrm{Gap}_{\mathrm{no\text{-}MRLB}}(100\mathrm{K})
\triangleq
\frac{\mathrm{FLOPs}_{\mathrm{STCA}}(100\mathrm{K})}{\mathbb{E}[\mathrm{FLOPs}]_{\mathrm{SeqO1}}(100\mathrm{K};p_{\mathrm{hit}}^{\mathrm{train}})}
\approx
\frac{5059.46}{108.10}
\approx 46.80\times.
\label{eq:gap_no_mrlb_100k}
\end{equation}

\paragraph{Training with MRLB.}
With reordering, STCA has no materialized per-layer $XW_K/XW_V$ to reuse; under MRLB only the per-layer history-side pre-FFNs are amortized:
\begin{equation}
\mathrm{FLOPs}^{\mathrm{MRLB}}_{\mathrm{STCA}}(L;R)
=
\frac{12N_{\mathrm{stca}}LD^2}{R}
+
N_{\mathrm{stca}}(4LDh+20D^2).
\end{equation}
For SequenceO1, we amortize all SA computation and the adapter, and in sketch-STCA we amortize only its per-layer pre-FFNs:
\begin{align}
\mathbb{E}[\mathrm{FLOPs}]^{\mathrm{MRLB}}_{\mathrm{SeqO1}}(L;R,p_{\mathrm{hit}})
&=
N_{\mathrm{stca}}(4L_sDh+20D^2)
+
\frac{\mathrm{FLOPs}_{\mathrm{adapt}}(L_s;d,D)+12N_{\mathrm{stca}}L_sD^2+(1-p_{\mathrm{hit}})\cdot \mathrm{FLOPs}_{\mathrm{SA}}(L;N_{\mathrm{sa}},L_s,d)}{R}.
\end{align}
At $L=100\mathrm{K}$, $R_{\mathrm{train}}=40$, and $p_{\mathrm{hit}}^{\mathrm{train}}=0.5$, we obtain
\[
\mathrm{STCA{+}MRLB}_{\mathrm{train}}\approx 152.13\ \mathrm{GFLOPs/target},\quad
\mathrm{SeqO1{+}MRLB}_{\mathrm{train}}\approx 3.046\ \mathrm{GFLOPs/target},
\]
so the expected training-side FLOP ratio becomes
\begin{equation}
\mathrm{Gap}_{\mathrm{train}}(100\mathrm{K})
\triangleq
\frac{\mathrm{FLOPs}^{\mathrm{MRLB}}_{\mathrm{STCA}}(100\mathrm{K};R_{\mathrm{train}})}
{\mathbb{E}[\mathrm{FLOPs}]^{\mathrm{MRLB}}_{\mathrm{SeqO1}}(100\mathrm{K};R_{\mathrm{train}},p_{\mathrm{hit}}^{\mathrm{train}})}
\approx
\frac{152.13}{3.046}
\approx 49.94\times.
\label{eq:gap_train_100k}
\end{equation}

\paragraph{Inference / serving-side amortization.}
Using the same formulas but with serving-side reuse ratio $R_{\mathrm{infer}}=300$ and cache hit rate $p_{\mathrm{hit}}^{\mathrm{infer}}=0.6$, we obtain
\[
\mathrm{STCA}_{\mathrm{infer}}\approx 43.076\ \mathrm{GFLOPs/target},\quad
\mathrm{SeqO1}_{\mathrm{infer}}\approx 0.67419\ \mathrm{GFLOPs/target},
\]
so the expected inference-side FLOP ratio becomes
\begin{equation}
\mathrm{Gap}_{\mathrm{infer}}(100\mathrm{K})
\triangleq
\frac{\mathrm{FLOPs}^{\mathrm{MRLB}}_{\mathrm{STCA}}(100\mathrm{K};R_{\mathrm{infer}})}
{\mathbb{E}[\mathrm{FLOPs}]^{\mathrm{MRLB}}_{\mathrm{SeqO1}}(100\mathrm{K};R_{\mathrm{infer}},p_{\mathrm{hit}}^{\mathrm{infer}})}
\approx
\frac{43.076}{0.67419}
\approx 63.89\times.
\label{eq:gap_infer_100k}
\end{equation}

\paragraph{Takeaway.}
Cross-target reuse increases the expected FLOP reduction at 100K relative to the setting without MRLB.
Under training-side MRLB with $R_{\mathrm{train}}=40$ and $p_{\mathrm{hit}}^{\mathrm{train}}=0.5$, the gap becomes $\sim\!49.94\times$;
under inference/serving-side reuse with $R_{\mathrm{infer}}=300$ and $p_{\mathrm{hit}}^{\mathrm{infer}}=0.6$, it further increases to $\sim\!63.89\times$.
This is because reordering leaves STCA with only the per-layer pre-FFNs as reusable components, while SequenceO1 can amortize the entire lightweight SA sketching stage and keep heavyweight computation concentrated in target-conditioned interaction over the fixed sketch length $L_s$.

\end{document}